\documentstyle[preprint,aps,epsfig]{revtex}

\newif\iftightenlines\tightenlinesfalse
\tightenlines\tightenlinestrue

\def\mz{M_Z} \def\mgut{M_{GUT}} \def\mpl{M_{Planck}} 
\def\eslt{\not\!\!{E_T}} \def\to{\rightarrow} \def\Phat{\hat{\Phi}}
\def\te{\tilde e}  \def\tt{\tilde t} \def\tu{\tilde u}
\def\tc{\tilde c} \def\ts{\tilde s} \def\tb{\tilde b} 
\def\td{\tilde d} \def\tQ{\widetilde Q}  
 \def\ttau{\tilde \tau} 
\def\tg{\tilde g}  
 \def\tw{\widetilde W} \def\tz{\widetilde Z}
 \def\CM{\cal M}

\begin{document}
\preprint{\vbox{\baselineskip=14pt%
   \rightline{IFUSP-1551/2002}
   \rightline{UH-511-1006-02}
}}

\title{Higgs-mediated leptonic decays of $B_s$ and $B_d$ mesons as
probes of supersymmetry}
\author{J.~K.~Mizukoshi$^1$, 
Xerxes Tata$^2$ and Yili Wang$^2$}
\address{$^1$Instituto de F\'{\i}sica, Universidade  de S\~ao Paulo,
C.P. 66318, 05315-970, S\~ao Paulo, Brazil.}

\address{
$^2$Department of Physics and Astronomy,
University of Hawaii,
Honolulu, HI 96822, USA.
}
\date{\today}
\maketitle
\begin{abstract}
If $\tan\beta$ is large, down-type quark mass matrices and Yukawa couplings
cannot be simultaneously diagonalized, and flavour violating
couplings of the neutral Higgs bosons are induced at the 1-loop
level. These couplings lead to Higgs-mediated contributions to the
decays $B_s \to \mu^+\mu^-$ and $B_d \to \tau^+\tau^-$, at a level that
might be of interest for the current Tevatron run, or possibly, at
$B$-factories. We evaluate the branching ratios for these decays
within the framework of minimal gravity-, gauge- and anomaly-mediated SUSY
breaking models, and also in $SU(5)$ supergravity models with non-universal
gaugino mass parameters at the GUT scale. We find that the
contribution from gluino loops, which seems to have
been left out in recent phenomenological analyses, is significant. We
explore how the branching fraction varies in these models, emphasizing
parameter regions consistent with other observations.
\end{abstract}

\newpage

\section{Introduction} \label{intro}

Supersymmetry provides a promising way to stabilize the electroweak
scale if superpartners are lighter than ${\cal O}(1)$~TeV \cite{stable}. All
SUSY theories \cite{rev} necessarily contain many scalar fields,
resulting in multiple potential sources of flavour violation. Indeed, in
constructing phenomenologically viable models, care has to be taken to
make sure that flavour violation is sufficiently suppressed. Within the
framework of the Minimal Supersymmetric Standard Model (MSSM) with
conserved $R$-parity, this is ensured (at tree level) by requiring that
the matter supermultiplets with weak isospin $T_3=1/2$ couple only to the
Higgs superfield ${\hat h}_u$, while those with $T_3=-1/2$ couple just
to the Higgs superfield ${\hat h}_d$. 

At the one loop level, however, a coupling of ${\hat h}_u$ to down type
fermions is induced.  This induced coupling leads to a new contribution
\cite{hrs}, proportional to $\langle h_u \rangle$, to the down type
fermion mass matrix. Although this contribution is suppressed by a loop
factor relative to the tree-level contribution, this suppression is
partially compensated if the ratio ${\langle h_u \rangle \over \langle
h_d \rangle} \equiv \tan\beta$ is sufficiently large.  As a result, down
type Yukawa coupling matrices and down type quark mass matrices are no longer
diagonalized by the same transformation, and flavour violating couplings
of neutral Higgs scalars $h, H$ and $A$ emerge. Of course, in the limit of
large $m_A$, the Higgs sector becomes equivalent to the Standard Model
(SM) Higgs sector with a light Higgs boson $h \simeq h_{SM}$, and the
effects of flavour violation decouple from the low energy theory .  The
interesting feature is that the flavour-violating couplings of $h$, $H$
and $A$, {\it do not decouple for large superparticle mass parameters}:
being dimensionless, these couplings depend only on ratios of these mass
parameters\footnote{These not only include sparticle masses, but also
the superpotential parameter $\mu$ and also the soft SUSY breaking
$A$-parameters.}, and so, remain finite even for very large values of
SUSY mass parameters. 

As pointed out by many authors
\cite{gaur,babu,chank,old,bobeth,isidori,dedes,arnowitt,bobeth2,buras,baek},
this flavour violating neutral Higgs boson coupling results in a
potentially observable branching fraction for the decay $B_s \to
\mu^+\mu^-$ mediated by the neutral Higgs bosons, $h$, $H$ and $A$.  If
$\tan\beta$ is sufficiently large the most important contribution to the
amplitude for this decay then scales as $\tan^3\beta$: two of these
factors arise via the down type quark and lepton Yukawa couplings which
are each $\propto 1/\cos\beta \simeq \tan\beta$, while the last factor
arises because the offset between the down quark Yukawa coupling
matrices and the mass matrices discussed above also increases with
$\tan\beta$.

Within the SM, the branching fraction for this decay is $\sim
3.4\times 10^{-9}$~ \cite{sm}. This is very far from its current
experimental upper limit of $2.6
\times 10^{-6}$ obtained by the CDF collaboration\cite{cdf}.
Within the SUSY framework, however, this
branching fraction can be enhanced by a very large factor and the SUSY
contribution may dominate its SM counterpart if $\tan\beta$ is sufficiently
large. Indeed, the CDF bound on this branching
fraction already constrains some portions of the SUSY parameter space
\cite{arnowitt,dedes,sugra}. It is, therefore, reasonable to expect that
the data from the Fermilab Main Injector will probe significant ranges
of SUSY parameters. Moreover, because the SUSY amplitude does not
decouple for large sparticle masses, it is possible that the CDF or D\O
\ experiments may detect new physics via a measurement of
$B(B_s \to \mu^+\mu^-)$
 even without direct detection of sparticles at the Tevatron.

Many of the analyses\cite{chank,old,bobeth,isidori,bobeth2,buras} of $B_s \to
\mu^+\mu^-$ have been performed within the MSSM framework. Since
sparticle mass matrices as well as soft SUSY breaking scalar coupling
matrices both include intrinsic sources of flavour violation, strictly
speaking the framework is too broad to be predictive.\footnote{It would,
of course, be possible to obtain constraints on flavour-mixing
elements of the squark mass matrix or $A$-parameters.} Even with
assumptions that set tree-level flavour violating effects to zero, the
model still has too many free parameters to correlate the branching
fraction for $B_s \to \mu^+\mu^-$ decay with other observables.

This situation is in sharp contrast to that in highly predictive
scenarios such as the minimal supergravity (mSUGRA) model \cite{sug},
the gauge-mediated SUSY breaking (GMSB) model \cite{gmsb}, or the
minimal anomaly-mediated SUSY breaking (mAMSB) model \cite{amsb} in
which much of SUSY phenomenology is analysed these days. In each of
these scenarios, SUSY is assumed to be dynamically broken in a ``hidden
sector'', comprised of fields that interact with SM particles (and their
superpartners) only
via gravity. It is the mechanism by which SUSY breaking is communicated to
the superpartners of SM particles that distinguishes these frameworks
from one another.  These models are all completely specified by just a
handful of parameters, usually defined at a scale, much higher than the
weak scale, where the physics is simple.

The main goal of this paper is to examine the prospects for observing
the decay $B_s \to \mu^+\mu^-$ in Tevatron experiments within these
framework of well-studied supersymmetry models. We also consider
gravity-mediated SUSY breaking models based on $SU(5)$
where the order parameter for SUSY breaking also breaks the GUT
symmetry\cite{anderson}, leading to non-universal GUT scale gaugino mass
parameters\cite{earliernonuni}.  With an integrated luminosity of
2~fb$^{-1}$, experiments at the Tevatron are expected to be sensitive to
a branching fraction below $\sim 10^{-7}$. With a still bigger data
sample (that is expected to accummulate before the Large Hadron Collider (LHC)
begins operation) the sensitivity should be even greater.

Of course, while detection of a signal would point to new physics, it
would not necessarily establish this new physics to be
supersymmetry. Within these specific scenarios, however, it is possible
to predict other signals that might also be simultaneously present if an
observed $B_s \to \mu^+\mu^-$ signal is to be attributed to any
particular supersymmetric framework. The first examination of the
Higgs-mediated $B_s \to \mu^+\mu^-$ decay in the supersymmetric context
that we know about was performed within the mSUGRA
framework~\cite{gaur}. However, it is only in the past year that the
observability of this signal as a function of mSUGRA parameters been
systematically investigated \cite{dedes,arnowitt}. A 
corresponding study\cite{baek} of
this signal within the gauge-mediated and the anomaly-mediated SUSY
breaking scenarios also appeared while this paper was in preparation. As
far as we are aware, there is no study of this decay in models with
non-universal gaugino masses.  As discussed in the next section we have
improved upon the existing phenomenological analyses, which consider
flavour violating effects just from chargino loops, in that we also
include effects from gluino loops which we find to be substantial.

The Higgs-mediated mechanism for the $\mu$-pair decay of $B_s$ mesons
also allows the decay $B_s \to \tau^+\tau^-$. In fact, since the
leptons couple via their Yukawa interactions, the branching ratio for
the latter decays would be expected to be enhanced by
$(m_{\tau}/m_{\mu})^2$.  Unfortunately, identification of $\tau$ pair decays of
$B_s$ would be very difficult in Tevatron experiments because their
invariant mass cannot be accurately reconstructed. Moreover, $B_s$ pair
production is not kinematically accessible at the $B$-factories
currently in operation at SLAC or KEK. The BELLE and BABAR experiments
at these facilities have already accumulated $\sim 70$M $B_d$ mesons,
and this sample is expected to increase to $\sim 500-1000$M $B_d$s in a
few years.  These considerations motivated us to also examine the
branching fraction for $B_d \to \tau^+\tau^-$. Although this decay is
suppressed by Kobayashi-Maskawa (KM) matrix elements relative to
corresponding decays of $B_s$, this is offset by the larger $\tau$
Yukawa coupling to the Higgs bosons. We are not aware of any studies of
the sensitivity of BELLE or BABAR to $B_d \to \tau^+\tau^-$ decays. We
will, therefore, confine ourselves to mapping out the branching fraction
for this decay without making any representation of its detectability in
experiments at $B$-factories.

The same flavor-violating couplings of neutral Higgs bosons that we have
discussed can also lead to significant contributions to other processes,
such as the exclusive decays $B_s \to K\ell^+\ell^-$ and $B_s
\to\ell^+\ell^-\gamma$, the semi-inclusive decay $B_s \to
X_s\ell^+\ell^-$, and to $\Delta F=2$ processes like $B_{d,s}$ and
$K$-meson mixing \cite{otherdec} that may be probed by experiment.
Their leptonic cousins lead to $\tau\to \mu\mu\mu$ and $\mu \to eee$
decays with branching fractions that might potentially be within reach
of future experiments\cite{bk2}.

The remainder of this paper is organized as follows. In
Sec.~\ref{formalism} we describe our computation of the amplitude for
the decay, focussing on the improvements that we have made, and on the
approximations that we have used to simplify the
analysis. Sec.~\ref{results} contains our main results. Here we analyse
the branching fractions for the decays $B_s \to \mu^+\mu^-$ and $B_d \to
\tau^+\tau^-$ within the mSUGRA, mGMSB and mAMSB models, as well as in
$SU(5)$ supergravity models with non-universal GUT scale gaugino
masses. We also comment on other observables (with emphasis on the other
flavour-violating decay $b \to s\gamma$) for parameter regions where
$B(B_s \to \mu^+\mu^-)$ may be observable at the Tevatron. We conclude
in Sec.~\ref{summary} with a summary of our results.

\section{Higgs-Mediated Leptonic Decays of ${\bf B_{\lowercase{s,d}}}$ Mesons }\label{formalism}

At tree level, the down (up) type quarks couple to the Higgs field
$h_d^0$ ($h_u^0$) via the Yukawa coupling matrix ${\bf f_d}$ ($ {\bf
f_u}$). 
Flavour violation in Higgs field interactions occurs because couplings
of the field $h_u^0$ to down type quarks is induced at the one loop
level. There are two distinct sets of SUSY sources of this coupling at
the 1-loop level \cite{babu}.  Down type quarks can couple to
$h_u^0$ via 
\begin{enumerate}
\item a chargino up-type squark loop, with the Higgs field
attaching to the squarks via the ${\bf Af_u}$ elements of the soft SUSY
breaking trilinear scalar coupling matrix, or 

\item via a gluino down-type squark loop, with the Higgs field attaching
to the squark via supersymmetric interactions proportional to the matrix
$\mu{\bf f_d}$. We ignore analogous contributions from neutralino loops;
see however, Ref.\cite{bobeth2}. 
\end{enumerate}

The gluino loop, at first sight, appears to give no flavour
violation. In the flavour basis, gluino interactions are flavour
diagonal, so that the Higgs-squark-squark coupling is the only source of
flavour violation. Since the flavour structure of this coupling is
proportional to the down quark Yukawa matrix ${\bf f_d}$ ({\it i.e.} the
same as the tree-level Yukawa matrix) it does not lead to flavour
violation. This reasoning is incorrect because the down-squark matrices
need not be flavour diagonal. In other words, the gluino contribution to
flavour violation {\it depends on the a priori unspecified structure of
the down squark mass matrices.} The flavour-violating coupling from the
chargino contribution above depends on the {\it essentially unknown}
matrix ${\bf Af_u}$.  These considerations make it clear why it is
difficult to make predictions of the flavour violating branching
fractions without resort to specific models, although as we said it is
possible to constrain certain (flavour-mixing) matrix elements in a more
general analysis. We will, in the following, focus on the four
models introduced in Sec.~\ref{intro}. The reader should, however,
keep in mind that our results for the branching fraction are specific to
these models, and that simple modifications to the flavour structure in
the sparticle sector could lead to a very different answer.

It is instructive to analyse the flavour structure of the gluino-induced
flavour violation. Following Ref.~\cite{babu}, we will work in the quark
basis where the up type Yukawa coupling matrix is diagonal, and the down
quark Yukawa coupling matrix has the form, ${\bf f_d} = {\bf
DV^\dagger_{KM}}$, where ${\bf D}$ is the matrix obtained by
diagonalizing ${\bf f_d}$ and ${\bf V_{KM}}$ is the Kobayashi-Maskawa
matrix (uncorrected for SUSY contributions). In the mass basis
$[d'_{Li},d'_{Rj}]$ for quarks, the flavour structure of the gluino
contribution can be written as,
\begin{equation}
(Z_R)_{jl}(Z_R^\dagger {\bf D} Z_L)_{lk}(Z_L)_{ki}C_0(m_{\td_{Lk}}^2,
m_{\td_{Rl}}^2, m_{\tg}^2),
\label{eq:gluino}
\end{equation}
where $m_{\td_{Lk}}^2$ ($m_{\td_{Rl}}^2$) is the $k$th ($l$th) eigenvalue of
the squared mass matrix for left (right) type squarks, $Z_L$ ($Z_R$)
is a unitary matrix that transforms the superpartners of $d'_L$ ($d'_R$)
to the basis in which the left (right) down squark squared mass matrix is
diagonal and $C_0$ is a loop function defined by,
\begin{equation}
C_0(x,y,z) = \frac{1}{y^2-z^2}\biggl[\frac{y^2}{x^2-y^2}\ln\biggl(
\frac{y^2}{x^2}\biggr) - \frac{z^2}{x^2-z^2}\ln\biggl(\frac{z^2}{x^2}\biggr)
\biggr]\;. 
\label{eq:C0}
\end{equation}
The following features are worth noting:
\begin{itemize}
\item If the different types of
left and right down squarks have common masses $m_L$ and $m_R$ (not
necessarily equal), the function $C_0$ will be independent of 
$k$ and $l$ and the flavour violating contribution from
(\ref{eq:gluino}) vanishes. 

\item In a wide class of models the left (and, separately, the right)
squark mass squared matrices are proportional to the unit matrix at some
scale, and only Yukawa interactions distinguish between
flavours.\footnote{Operationally, this means that the squark mass
matrices have the form, ${\bf m_{\td_L}^2}= \bar{m}_L^2\left[1+c_L{\bf
f_d^\dagger f_d} + c'_L {\bf f_u^\dagger f_u }\right]$ and ${\bf
m_{\td_R}^2}= \bar{m}_R^2\left[1+c_R{\bf f_d f_d^\dagger}\right]$, with
$c_L, c'_L$ and $c_R$ as constants.}  Then, the matrix ${\bf
m_{\td_R}^2}$ is diagonal in the basis of superpartners of $d'_R$, and
$Z_R =I$. This may be seen by noting that the 1-loop renormalization
group equation for the singlet down squark mass matrix depends only on
the down type Yukawa coupling matrix.  The same is not the case for the
doublet squarks since the renormalization group equation now depends
also on the up type Yukawa couplings. In all such models, gluino-induced
flavour violation depends only upon the mass splittings in the left
squark sector. Note that all the models that we
consider all fall in this category.
\end{itemize}

\subsection{Flavour-violating Higgs boson interactions}
In our analysis we follow the approach in Ref. \cite{babu} and develop
an effective Lagrangian to describe the induced flavour violation in the
neutral Higgs boson sector.  We
will assume that squarks with the same quantum numbers have a common
mass at some energy scale, and further that superpotential Yukawa
interactions are the sole source of flavour violation in the remainder
of this analysis. We take the $SU(3)$ gaugino mass parameter to be
positive by convention: other gaugino mass parameters may have either sign.
As in Ref. \cite{babu}, we work in the basis where
${\bf f_u}$ is diagonal and neglect masses for $d$ and $s$ quarks. 
Neglecting terms that are second order in the Wolfenstein
parameter $\lambda$, the effective couplings of down quarks to the Higgs
fields of the MSSM can be written as,\footnote{This is the analogue of
Eq.~(7) of Ref.\cite{babu}.}

\begin{equation}
-{\cal L}_{eff} = \overline{D}_R {\bf f_D}Q_L h_d +
\overline{D}_R {\bf f_D}[a_g {\bf {\CM}}^g + 
a_u {\bf {\CM}}^u {\bf f_U^\dagger f_U} + a_w {\bf {\CM}}^w] Q_Lh_u^* +h.c.\;,
\label{eq:lageff}
\end{equation} 
with
\begin{equation}
a_g =  -\frac{2\alpha_s}{3\pi} \mu m_{\tg}, \;\;\;
a_u = -\frac{1}{16\pi^2} \mu A_t, \;\;\;
a_w = \frac{g^2}{16\pi^2} \mu M_2\;,
\end{equation}
and
\begin{eqnarray}
{\bf \CM}^g  &=& \mbox{diag} (C^{g, i}_{0, 1}, C^{g, i}_{0, 2}, 
C^{g, i}_{0, 3}),
\;\; C^{g, i}_{0, j} \equiv  C_0(m_{\tg}, m_{\td_{i, R}}, m_{\td_{j, L}})\;,
\nonumber \\
{\bf \CM}^u  &=& \mbox{diag} (0, 0, C^u_{0, 3}), 
\;\; C^u_{0, 3} \equiv  C_0(\mu, m_{\tt_R}, m_{\tt_L})\;,
\label{eq:c0}
\\
{\bf \CM}^w  &=& \mbox{diag} (C^w_{0, 1}, C^w_{0, 2}, C^w_{0, 3}),
\;\; C^w_{0, j} \equiv  C_0(M_2, \mu, m_{\tu_{j, L}})\;.
\nonumber
\end{eqnarray}
The loop function $C_0(x,y,z)$ is given in Eq.~(\ref{eq:C0}).
Note that the right squark index in the matrix ${\bf \CM}^g$ in
(\ref{eq:lageff}) is fixed by the corresponding quark; matrix
multiplication is implied with
the left squark index of the ``diagonal matrix'' ${\bf \CM}^g$.

The second term of (\ref{eq:lageff}) contains the loop-induced coupling
of $h_u^0$ to down quarks. The first entry in this term comes from
the gluino loop, while the last two entries arise due to chargino
loops. We see immediately that if $\tQ_{j,L}$ have a common mass, there
is no flavour violation from the gluino term since it then has exactly
the same flavour structure as the tree-level term. The term depending on
${\bf \CM}^u$ is proportional to the squared top Yukawa coupling, and
arises from the Higgsino components of the chargino, with the fermion
chirality flip proportional to the Higgsino mass $\mu$. The last term in
(\ref{eq:lageff}), which comes from gaugino-higgsino mixing and again
vanishes if left type squarks have a common mass. It is, generally
speaking, much smaller than the other two terms. 

The flavour changing coupling between second and third generation down
quarks is given by,
\begin{eqnarray}
{\cal L}_{FCNC} &=& \frac{\bar{f}_b}{\sin\beta}V_{tb}^\ast V_{ts}
\chi_{FC} \bar{b}_R s_L (\cos\beta h_u^\ast - \sin\beta h_d)
\nonumber \\
&=& \frac{\bar{f}_b}{\sqrt{2}\sin\beta}V_{tb}^\ast V_{ts}
\chi_{FC} \bar{b}_R s_L [\cos(\beta+\alpha)h - \sin(\beta+\alpha)H -iA]\;,
\label{eq:efflag}
\end{eqnarray}
where $\alpha$ is the mixing angle in the neutral scalar Higgs boson
sector, $\bar{f}_b$ is the 
``physical $b$ Yukawa coupling'' defined by\cite{babu},
\begin{equation}
\bar{f}_b = f_b [1+(a_g C^{g, 3}_{0, 3}  + 
a_w C^w_{0,3} + a_u C^u_{0, 3} f_t^2)\tan\beta]\;,
\label{eq:physyuk}
\end{equation}
and
\begin{equation}
\chi_{FC} = -\frac{[a_g (C^{g,3}_{0,3} - C^{g,3}_{0,2}) 
 + a_w(C^w_{0,3} - C^w_{0,2}) +  a_u C^u_{0,3} f_t^2]\tan\beta}
{[1+(a_g C^{g,3}_{0,2} + a_w C^w_{0,2})\tan\beta]
[1+(a_g C^{g,3}_{0,3}  + a_w C^w_{0,3} + a_u C^u_{0,3} f_t^2)\tan\beta]}\;.
\label{eq:chi}
\end{equation}
We have checked that if we
multiply (\ref{eq:physyuk}) 
by $v_d$, we recover the SUSY correction to the $b$ quark 
mass as given by Pierce {\it et al.} \cite{pierce}.
For degenerate squarks, we recover the expression for $\chi_{FC}$ in 
Ref. \cite{babu} by setting $a_w=0$ in the 
denominator of (\ref{eq:chi}).


The flavour violating couplings between first and third generation down
squarks are obtained by obvious substitutions.
If squarks of the first and second generations are 
degenerate (as is the case in many models, including all models we
consider in this paper) we can replace $s_L$ by  $d_L$ without changing $\chi_{FC}$. 
Therefore, for the $B_d$ calculation we only have to replace 
$V_{ts} \to V_{td}$ in Eq.~(\ref{eq:efflag}), keeping everything else 
the same.

Eq.~(\ref{eq:chi}) ignores intra-generation squark mixing. Assuming that
this is significant only for third generation squarks, we find the
modified result,
\begin{equation}
\chi_{FC} = -\frac{N}{D} \tan\beta\;,
\end{equation}
with
\begin{eqnarray}
N &=& a_g \biggl[C_0(m_{\tg}, m_{\tb_1}, m_{\tb_2}) - 
s^2_b C_0(m_{\tg}, m_{\tb_1}, m_{\ts_L}) -
c^2_b C_0(m_{\tg}, m_{\tb_2}, m_{\ts_L})\biggr] +
a_u f^2_t C_0(\mu, m_{\tt_1}, m_{\tt_2})
\nonumber \\ 
&& + a_w \biggl[c^2_t C_0(M_2, \mu, m_{\tt_1}) + 
s^2_t C_0(M_2, \mu, m_{\tt_2}) - C_0(M_2, \mu, m_{\tc_L})\biggr]\;,
\nonumber \\
D &=& \bigg\{1 + a_g \tan\beta
\biggl[s^2_b C_0(m_{\tg}, m_{\tb_1}, m_{\ts_L}) -
c^2_b C_0(m_{\tg}, m_{\tb_2}, m_{\ts_L})\biggr]  +
a_w \tan\beta C_0(M_2, \mu, m_{\tc_L})\biggr\} 
\nonumber \\
&& \times \biggl\{1 + a_g \tan\beta C_0(m_{\tg}, m_{\tb_1}, m_{\tb_2}) +
a_w \tan\beta \biggl[ c^2_t C_0(M_2, \mu, m_{\tt_1}) + 
s^2_t C_0(M_2, \mu, m_{\tt_2})\biggr] 
\nonumber \\
&&+ a_u  f^2_t \tan\beta 
C_0(\mu, m_{\tt_1}, m_{\tt_2}) \biggr\}\;,
\label{eq:chif}
\end{eqnarray}
where $s_{b,t} \equiv \sin\theta_{b,t}, c_{b,t} \equiv 
\cos\theta_{b,t}$, and $\theta_{b,t}$ are the bottom and top squark
mixing angles, respectively.

We should note that our calculation includes only terms
that are most enhanced by powers of $\tan\beta$. Our calculation is,
therefore, valid only when $\tan\beta \agt 25-30$. 
We will subsequently see that for smaller values of
$\tan\beta$ the branching fraction of interest is too low to be of
interest at the Tevatron, so that this is not a serious handicap for
the purpose of our analysis.
We remark that in the loop
functions we have assumed that the chargino masses are well approximated by
$|M_2|$ and $|\mu|$. 

Finally, the leptonic couplings of the neutral Higgs bosons $h$, $H$ and
$A$ of the MSSM that are
needed to complete the evaluation of Higgs boson mediated leptonic
decays of $B_s$ or $B_d$ are given by,
\begin{equation}
{\cal L}_{H\ell\bar{\ell}} = -{{gm_{\ell}} \over {2M_W\cos\beta}}\left[h\sin\alpha +
H\cos\alpha \right]\bar{\ell}\ell + {{igm_{\ell}\tan\beta} \over {2M_W}}
\bar{\ell}\gamma_5\ell A \;.
\label{eq:laglep}
\end{equation}

\subsection{The branching ratio for $B_{\lowercase {s,d}} \to \ell^+
\ell^-$ decays}
The effective
Hamiltonian \cite{sm} for $B_{d'} \to \ell^+ \ell^-, d'=s,d$, is given
by
\begin{displaymath}
H = \frac{G_F}{\sqrt{2}}V_{td'}^\ast V_{tb}[c_{10} {\cal O}_{10} + c_{Q_1}
Q_1 + c_{Q_2} Q_2] + h.c. \;,
\end{displaymath}
with the relevant operators being given by,
\begin{equation}
{\cal O}_{10} = \frac{e^2}{4\pi^2} \bar{d'}_L \gamma^\mu b_L \bar{\ell} 
\gamma_\mu \gamma_5 \ell, \;\;\;
Q_1 = -\frac{e^2}{4\pi^2} \bar{d'}_L b_R \bar{\ell} \ell, \;\;\;
Q_2 = -\frac{e^2}{4\pi^2} \bar{d'}_L b_R \bar{\ell} \gamma_5 \ell \;.
\end{equation}
The coefficient $c_{10}$ that encapsulates the SM contribution is given
by \cite{sm},
\begin{equation}
c_{10} = -\frac{Y(x_t)}{s_W^2} \approx -4.2\;.
\end{equation}
The function $Y(x_t)$ that appears here is defined in Ref.\cite{sm}. 
The SUSY contributions can be obtained by
applying the matching condition between the amplitude given by the effective 
Hamiltonian and the one obtained using (\ref{eq:efflag}) and
(\ref{eq:laglep}). We find,
\begin{eqnarray}
c_{Q_1} &=& \frac{2\pi}{\alpha} \chi_{FC} \frac{m_b m_\ell}{\cos^2\beta
\sin^2\beta}\biggl(\frac{\cos(\beta+\alpha)\sin\alpha}{m^2_h} -
\frac{\sin(\beta+\alpha)\cos\alpha}{m^2_H}\biggr)\;, \nonumber
\\
c_{Q_2} &=& \frac{2\pi}{\alpha} \chi_{FC} \frac{m_b m_\ell}{\cos^2\beta}
\frac{1}{m^2_A}\;.
\label{eq:CQ}
\end{eqnarray}
Note that as $m_A \to \infty$, $\alpha+\beta \to {\frac{\pi}{2}}$ and 
$m_H \simeq m_A$. This then implies $c_{Q_1} \simeq -c_{Q_2}$, an
observation that will prove useful later.

Finally, using the hadronic matrix elements,
\begin{eqnarray*}
\langle 0|\bar{b}\gamma^\mu \gamma_5 d'(x)|B_{d'}(P)\rangle &=&
 i f_{B_{d'}} P^\mu e^{-iP \cdot x}\;,
\\
\langle 0|\bar{b} \gamma_5 d'(x)|B_{d'}(P)\rangle &=&
- i f_{B_{d'}} \frac{m^2_{B_{d'}}}{m_b+m_{d'}} e^{-iP \cdot x}\;,
\end{eqnarray*}
we can write the 
branching ratio for the decay $B_{d'} \to \ell^+ \ell^-$ as,
\begin{eqnarray}
B(B_{d'} \to \ell^+ \ell^-) &=& \frac{G_F^2\alpha^2 m^3_{B_{d'}} 
\tau_{B_{d'}} f^2_{B_{d'}}}{64\pi^3}
|V^\ast_{tb}V_{td'}|^2\sqrt{1-\frac{4m^2_\ell}{m^2_{B_{d'}}}}
\nonumber \\
&&\times\biggl[\biggl(1-\frac{4m^2_\ell}{m^2_{B_{d'}}}\biggr)
\bigg|\frac{m_{B_{d'}}}
{m_b+m_{d'}}c_{Q_1}\bigg|^2 + \bigg|\frac{2m_\ell}{m_{B_{d'}}}c_{10}
- \frac{m_{B_{d'}}}{m_b+m_{d'}} c_{Q_2}\bigg|^2\biggr]
\label{eq:width}
\end{eqnarray}

\subsection{Comparision with other studies}

A complete calculation of the 1-loop chargino-induced flavour violating
MSSM Higgs boson couplings, along with the corresponding result for the
coefficients $C_{Q_1}$ and $C_{Q_2}$ in (\ref{eq:CQ}), may be found in
Ref.\cite{bobeth}. In comparision, we have only retained leading terms
in $\tan\beta$, so that our simplified calculation is valid only if
$\tan\beta \agt 25-30$. We have checked that our result for the
contribution to the coefficients $C_{Q_{1,2}}$ from the chargino graphs
indeed reduces to (5.13) of Ref.\cite{bobeth}, with
$m_{\tw_{1,2}}=|M_2|$ and $|\mu|$.  The results of Ref.\cite{bobeth}
have been used for the recent phenomenological
analysis\cite{dedes,arnowitt} within the mSUGRA framework.\footnote{The
earlier studies\cite{old,chank,isidori} were performed within the MSSM for
particular choices of parameters.} Our calculations improve upon these
in that we include contributions from the gluino-mediated flavour
violating contributions (which we will see are significant) first
discussed by Babu and Kolda\cite{babu}. We found it difficult to tell
whether or not these contributions were included in the recent
phenomenologial analysis of the mGMSB and mAMSB scenarios by Baek 
{\it et al.} \cite{baek}: they do not give any formulae nor do they discuss how
their calculation was performed.

While we were preparing this paper, Ref.\cite{bobeth2} appeared. This
study includes a complete calculation of gluino and also analogous
neutralino-induced contributions to flavour violating Higgs boson
couplings, but again within the framework of the MSSM with
specific assumptions about sfermion mass matrices. It is pointed out
that it is possible to find special regions of MSSM parameter space
where due to large cancellations between gluino and chargino
contributions, neutralino diagrams (which are usually small) can no
longer be neglected. Indeed, in this case even the ratio of the
branching fractions $B(B_d \to \mu^+\mu^-)/B(B_s \to \mu^+\mu^-)$ may be
different from that expected from ratios of Kobayashi-Maskawa matrix
elements.

\section {${\bf B(B_{\lowercase{s,d}} \to \ell^+\ell^-)}$ in SUSY Models: Results}\label{results}

We are now ready to proceed with the evaluation of $B(B_{s,d} \to
\ell^+\ell^-)$ in the various SUSY models discussed above.  For a given
point in the parameter space of these models, we use the program ISAJET
v 7.63 \cite{isajet} to evaluate the corresponding MSSM parameters that
we need for all phenomenological analysis, including the computation of
this branching ratio. This then allows us to assess the ranges of model
parameters where signals from $B(B_{s,d} \to \ell^+\ell^-)$ decays may
provide evidence of deviation from the SM. A clear advantage of this
approach is that for any {\it given model}, it is also possible to
compute SUSY contributions to other observables; {\it e.g.} $B(b \to
s\gamma)$ or $g_{\mu}-2$ that have already been constrained by
experiment\cite{bsgexp,g-2exp}. Moreover, constraints from the
non-observation of any sparticles \cite{spartexp} or Higgs bosons
\cite{higgsexp} which yield lower limits, $m_{\tw_1}\ge 103$~GeV,
$m_{\te} \ge 100$~GeV, $m_{\ttau_1} \ge 76$~GeV, $m_h \ge 113$~GeV and
$m_A \ge 100$~GeV, on their masses can be readily
incorporated.\footnote{Although the exact limits on sparticle masses
depend somewhat on the model, and also on where we are in parameter
space, the limits indicated here are applicable for a wide class of
models. The lower limit of 114~GeV on the mass of the SM Higgs boson has
to be translated into the limit on $m_h$. Except when $A$ is light, $h$
is essentially the SM Higgs boson and this limit applies essentially
without modification.}

ISAJET 7.63 includes several improvements over previous versions. From
our point of view, the most important of these is the improvement of the
bottom Yukawa coupling that enters the computation of $m_A$. We see from
(\ref{eq:CQ}) and (\ref{eq:width}) that the value of $m_A$ plays an
important role in the determination of the branching ratio of
interest. There are regions of parameter space where, because of
cancellations, $m_A$ may be considerably smaller than other sparticle
masses. Especially for these parameter ranges, the improved computation
of the Yukawa coupling plays a crucial role. 
 
The numerical values of various meson masses, lifetimes, decay constants
and Kobayashi-Maskawa mixing matrix elements that we use as inputs to
our analysis are,

$B_s$ mesons:
\[ m_{B_s} = 5.3696 \;\mbox{GeV}, \;\;\; f_{B_s} = 0.250 \;\mbox{GeV},\;\;\;
\tau_{B_s} = 1.493 \;\mbox{ps} \]

$B_d$ mesons:
\[ m_{B_d} = 5.2794 \;\mbox{GeV}, \;\;\; f_{B_d} = 0.208 \;\mbox{GeV},\;\;\;
\tau_{B_d} = 1.548 \;\mbox{ps} \]

Kobayashi-Maskawa matrix elements:
\[|V_{tb}| = 0.999,\;\;\; |V_{ts}| = 0.039,\;\;\; |V_{td}| = 0.009 \]

\subsection{Minimal supergravity model (mSUGRA)}\label{msugrares}

SUSY is assumed to be broken in a hidden sector consisting of fields
that interact with usual particles and their superparners only via
gravity.  SUSY breaking is communicated to the visible sector via
gravitational interactions, and soft SUSY breaking sparticle masses and
couplings are generated.  Without further assumptions, the scalar masses
can be arbitrary leading to flavor changing processes in conflict with
experiment.

Within the mSUGRA grand unified framework\cite{sug}, it is assumed that
at some high scale (frequently taken to be $\sim M_{GUT}$) all scalar
fields have a common SUSY breaking mass $m_0$, all gauginos have a mass
$m_{1/2}$, and all soft SUSY breaking scalar trilinear couplings have a
common value $A_0$. 
Electroweak symmetry breaking is assumed to occur radiatively. This
fixes the magnitude superpotential parameter $\mu$. The soft SUSY
breaking bilinear Higgs boson mass parameter can be eliminated in favour
of $\tan\beta$, so that the model
is completely specified by the parameter set:
\begin{equation}
m_0,\ m_{1/2},\ A_0,\ \tan\beta,\ sign(\mu ) .
\end{equation}
The weak scale SUSY parameters that enter the computation of sparticle
masses and couplings required for phenomenological analyses can be
obtained via renormalization group evolution between the scale of
grand unification and the weak scale. 

We use the program ISAJET to evaluate these MSSM parameters, and via
these the various masses and mixing angles that enter the flavour
violating coupling $\chi_{FC}$ given by
Eq.~(\ref{eq:chif}). The required partial width can then be computed
using Eq.~(\ref{eq:width}). 

In Fig.~\ref{fig:sugtanbA} we show the dependence of $B(B_s \to
\mu^+\mu^-)$ and $B(B_d \to\tau^+\tau^-)$ on {\it a})~$\tan\beta$ for
$A_0=0$, and {\it b})~$A_0$ for $\tan\beta=40$. We fix
$m_0=m_{1/2}=300$~GeV, and illustrate the results for both positive
(solid) and negative (dashed) values of $\mu$.   
Values of $\tan\beta$ ($|A_0|$) larger than the
corresponding value denoted by squares on the curves are where a
sparticle or Higgs boson mass (in this case, it is always $m_h$) falls
below its experimental bound. 
The uppermost set of
the dashed and solid lines corresponds to the branching fraction for the
decay $b\to s\gamma$, the scale for which is given on the right hand
axis. Our purpose in showing this figure is to understand the behaviour
of the leptonic decays of $B_d$ and $B_s$, so that the fact that $B(b
\to s\gamma)$ appears to be outside the allowed range should not bother
the reader.
Several features are worth noting.
\begin{enumerate}
\item Since we have neglected contributions from neutralino loops
\cite{bobeth2}, the branching ratio for $B_d \to \tau^+\tau^-$ decays is
larger than that for $B_s \to \mu^+\mu^-$ decays by just a constant
factor fixed by SM parameters.

\item Focussing for the moment on the low $\tan\beta$ end of the $B(B_s \to
\mu^+\mu^-)$ curves, we see that the solid and dashed lines lie on
either side of the SM value, reflecting the fact that the sign of the
SUSY contribution flips with the sign of $\mu$, and its interference
with the SM contribution goes from destructive for $\mu < 0$ to
constructive for positive values of $\mu$. Of course, this becomes
irrelevant for very large values of $\tan\beta$ where the SUSY
contribution dominates. 

\item A striking feature in both frames is the very sharp rise in the
branching ratio for $\mu< 0$. Naively, we expect the SUSY contribution
to behave as $\tan^6\beta/m_A^4$.\footnote{Recall that the contribution
from $h$ exchange is very small as long as $h$ is a SM-like Higgs, and
further, that $m_H \sim m_A$ in the same limit.}  The dashed curves in
frame {\it a}) indeed roughly show this scaling behaviour as long as we
stay away from the lower range of $\tan\beta$ values where interference
with the SM amplitude is significant. But this is somewhat fortituous
because the dashed curves in frame {\it b}) rise much faster than would
be naively expected. This is because it is also very important to take
into account differences in other MSSM parameters, most notably $A_t$
and $\mu$ that enter via $a_u$ and $a_g$ in (\ref{eq:lageff}). In
general, the dashed curves are steeper than the solid curves because,
for $\mu < 0$, $m_A$ decreases sharply as $\tan\beta$ increases: there
is no corresponding decrease of positive values of $\mu$.

\item It is interesting to see that both $B(B_s \to \mu^+\mu^-)$ and
$B(B_d \to \tau^+\tau^-)$ can be potentially very large (and even exceed the
current upper limit) for values of parameters where there are no {\it
direct} signals for SUSY or Higgs bosons. This region of parameters is
also excluded by the experimental value of the inclusive branching ratio
$B(b \to s\gamma)$ \cite{bsgexp}. It should be understood though that our
purpose in showing this is pedagogical. 

\item We have also examined the case with $m_0=1$~TeV and other
parameters as before, except that the allowed range of $A_0$ is much
larger. For positive values of $\mu$ and $A_0=0$, $B(B_s\to \mu^+\mu^-)$
varies between $(4-100)\times 10^{-9}$. For $\mu < 0$, this branching
fraction is close to or smaller than the SM prediction, except when
$46\alt \tan\beta\alt 48$ when the branching fraction shoots up by over
two orders of magnitide. For $\tan\beta=40$ as in Fig.~\ref{fig:sugtanbA}{\it
b} and $\mu> 0$, the corresponding dependence on $A_0$ is qualitatively
similar to that the figure. For negative values of $\mu$ the branching
fraction is close to its SM value for $|A_0| \alt 1.2m_0$, but rapidly
shoots up to beyond $10^{-7}$ when $|A_0|$ becomes large. As in the
$m_0=300$~GeV case, the rapid rise is driven by the rapid drop in $m_A$.

\end{enumerate}

Having used this pedagogical illustration to obtain an idea of how and
why the branching fractions vary with parameters, we now turn to an
exploration of these branching ratios in mSUGRA parameter space. Because
the sparticle spectrum is most sensitive to $m_0$ and $m_{1/2}$, the
$m_0-m_{1/2}$ plane provides a convenient way of displaying the results,
as illustrated in Fig.~\ref{fig:sugcont} for $A_0=0$ and {\it a})~$\mu >
0, \tan\beta=50$ and {\it b})~$\mu < 0, \tan\beta=45$.  The dark-shaded
regions are excluded by theoretical constraints: charge-breaking minima
or lack of electroweak symmetry breaking.  In the slant-hatched region,
$\tz_1$ is not the lightest supersymmetric particle (LSP). The region
covered by open circles is excluded by lower limits on sparticle masses
or on $m_A$. Below the dot-dashed line labelled $h$, $m_h$ is smaller
than 113~GeV: we show this separately because this limit is modified if
$m_A \alt 150$~GeV.  Our main results are the contours of $B(B_s \to
\mu^+\mu^-)$ (solid) and $B(B_d \to \tau^+\tau^-)$ (dashed) within the
mSUGRA framework.  The contours are labelled by the values of the
corresponding branching fractions. In frame {\it a}), the innermost
dashed contour corresponds to a branching fraction of $10^{-6}$.  From
frame {\it a}) we see that even in the region allowed by all these
constraints $B(B_s \to \mu^+\mu^-)$ ($B(B_d \to \tau^+\tau^-)$) may be
as large as $10^{-7}$ ($10^{-6}$), while from frame {\it b}) the
corresponding branching fraction may be considerably larger.  Also shown
are contours of fixed branching fraction for the decay $b \to s\gamma$
computed using the calculation of Ref.\cite{brhlik} with improvements
described in Ref.\cite{sugra}. Below the contours labelled $b \to
s\gamma$ in frame {\it a}) (frame {\it b})) the branching fraction is
smaller (larger) than 2$\times 10^{-4}$ ($5\times 10^{-4}$), and appear
to be comfortably outside our assessment $2.16\times 10^{-4} \alt B(b\to
s\gamma) \alt 4.34\times 10^{-4}$, \cite{sugra} for the allowed
experimental range\cite{bsgexp} of this branching fraction. We caution
the reader that the SUSY contribution, for large values of $\tan\beta$,
may have considerable theoretical uncertainty, so that this
``constraint'' should be interpreted with some care.\footnote{It is also
worth emphasizing that unlike constraints from direct searches,
constraints from $B(b\to s\gamma)$ are very sensitive to details of the
model. For instance, small amount of flavour mixing (from some unknown
physics) in the squark sector could lead to large differences in the
predictions for $b\to s\gamma$, and for that matter, $B_{q'} \to
\ell^+\ell^-$ decays. For this reason, we urge our experimental
colleagues to view theorists' assessment of ``excluded regions''
(especially when these are excluded due to SUSY loop effects as opposed
to direct searches) including those in this paper in the proper
perspective. It is logically possible that small deviations from the
defining assumptions of a particular framework such as mSUGRA may permit
much larger signals without being in conflict with current constraints.}

The dotted curves are contours of $B(B_s \to \mu^+\mu^-)=10^{-8}$,
obtained by calculating the branching ratio by retaining just the
chargino-loop in evaluating the SUSY contribution to the decay {\it
i.e.} if $a_g$ in (\ref{eq:lageff}) is set to zero. We remind the reader
that it is just this contribution has been included in recent
analyses\cite{dedes,arnowitt}.  We have checked that the dotted contour
in frame~{\it a}) is in good agreement with the corresponding contour in
Fig.~3 of Ref.\cite{dedes}. This provides quantitative confirmation of
the validity of our approximations, relative to the complete calculation
of Ref.\cite{bobeth}. More importantly, the difference between the
dotted contour and the corresponding solid contour highlights the
importance of retaining the effect of gluino loops whose contribution
appears to interfere destructively with the chargino contribution. Thus,
earlier conclusions\cite{dedes,arnowitt} about the SUSY reach of
Tevatron experiments may be over-optimistic.

To see how the sensitivity of these experiments varies with $\tan\beta$,
we show in Fig.~\ref{fig:sugtbcont} contours of $B(B_s \to
\mu^+\mu^-)=10^{-7}$ (solid lines) and $B(B_d \to \tau^+\tau^-)=10^{-6}$
(dashed lines) for the values of $\tan\beta$ that label the contours.
We take $A_0=0$ and show results for {\it a})~$\mu > 0$ and {\it
b})~$\mu < 0$.  The hatched region corresponds to $\tan\beta=35$. The
lines in frame {\it a}) terminate at the corresponding boundaries of the
theoretically forbidden regions. As expected, the region over which the
branching fraction may be probed at the Tevatron is very sensitive to
$\tan\beta$. Even for $\tan\beta=35$, this channel appears to yield a
better SUSY sensitivity than direct searches\cite{sugrarep} for an
integrated luminosity of $\sim 2$~fb$^{-1}$.

From Fig.~\ref{fig:sugcont} and Fig.~\ref{fig:sugtbcont} we note the
following:
\begin{itemize}
\item The branching fraction under study is significantly larger for
negative values of $\mu$. We have traced this to the structure of the
denominator of $\chi_{FC}$ in Eq.~(\ref{eq:chif}). Changing the sign of
$\mu$ changes the sign of all the $a_i$, so that a suppression for
positive $\mu$ changes to an enhancement for $\mu < 0$. Notice that
while our denominator factor reduces to that in Ref.\cite{babu} in the
appropriate limit, it differs from that in Ref.\cite{dedes} who do not
have the $a_g$ term in the second factor.\footnote{In Ref.\cite{bobeth},
which is a strict diagramatic calculation, does not have this
denominator. Dedes {\it et al.} who use the results of
Ref.\cite{bobeth}, include a denominator correction in their formulae.}
Quantitatively, this leads to differences of ${\cal O}$(10\%) and do not
affect the qualitative conclusions.

\item Although $\mu < 0$ is  generally thought to be disfavoured by the
determination of the muon anomolous magnetic moment by the E821
experiment\cite{g-2exp}, we advise caution in this regard: in contrast
to conventional wisdom \cite{usual} a
conservative estimate \cite{melnikov} of the theoretical error suggests
that there is a region allowed\cite{sugra} by this constraint, though
perhaps in conflict with $B(b \to s\gamma)$, where $B_s \to \mu^+\mu^-$
decays may provide the first hint of new physics if $\tan\beta \agt 42$
or so. This region would expand as Tevatron experiments accummulated
more data.

\item From these figures, we see that a sensitivity of $10^{-6}$ for
$B(B_d \to\tau^+\tau^-)$ would roughly yield the same reach as an order
of magnitude better sensitivity that is expected to be attained at the Tevatron
with an integrated luminosity of $\sim 2$~fb$^{-1}$. We do not know
whether experiments at $B$ factories will be able to attain this sensitivity. 
\end{itemize}

Up to now, in our scan of mSUGRA parameter space, we have only
considered models with $A_0=0$. To understand just how large the
branching ratio for $B_s \to \mu^+\mu^-$ can become for other values of
$A_0$, we have scanned mSUGRA models with $m_0$ between 100~GeV and
$\sqrt{50}m_{1/2}$ and $-3m_0 <A_0 < 3m_0$. In Fig.~\ref{fig:sugscan},
we show $B(B_s \to \mu^+\mu^-)$ as a function of $m_{1/2}$, for {\it
a})~$\tan\beta=55$, $\mu>0$, {\it b})~$\tan\beta=45$, $\mu < 0$, and
{\it c})~$\tan\beta=40$, $\mu < 0$. Models are accepted only if they
satisfy the theoretical and the experimental constraints from direct
searches for sparticles and Higgs bosons.  For each model, we
put a cross (dot) if $B(b \to s\gamma)$ lies within (outside)
$(2-5)\times 10^{-4}$.  We note the following:
\begin {enumerate}

\item Although we have shown just $B(B_s\to\mu^+\mu^-)$ in this figure,
the range of $B(B_d \to \tau^+\tau^-)$, which differs only by a constant 
factor can easily be estimated.

\item We see that for $\mu > 0$, in frame {\it a}) the branching
fraction is typically smaller than $10^{-8}$, and never exceeds $\sim
3\times 10^{-8}$ for models loosely satisfying the $b \to s\gamma $
constraint.  But for this constraint, it can be as high as
$10^{-7}$.

\item Turning to frame {\it b}) we see that for negative
values of $\mu$ the branching fraction can be as large as $10^{-5}$, but
most of these points are excluded by the experimental measurement of
$b\to s\gamma$ because the predicted branching fraction is too
large. For large values of $m_{1/2}$, however, $B(B_s \to \mu^+\mu^-)$
may even exceed $10^{-7}$ for parameter values allowed by $b\to s\gamma$
constraints. For the slightly smaller value of $\tan\beta$ shown in
frame {\it c}), we see that these large values of  
$B(B_s \to \mu^+\mu^-)$ are possible over a wide range of $m_{1/2}$. 

\item In all frames we see that although the branching ratio may be much
larger than the SM value for some set of SUSY parameters, it never falls
below about half the SM value. Experiments at the LHCb should be
sensitive (at the 3$\sigma$ level) down to this branching
fraction~\cite{lhc}, so that a non-observation of this decay in these
experiments would signal new physics other than the scenarios considered
here.\footnote{Admittedly we have not scanned all parameter
space. However, if $\tan\beta$ is small the SUSY contribution is
reduced, and we would expect the branching fraction to be closer to its
SM value.}

\item A striking feature of the figure is the ``gap'' for small
$m_{1/2}$ values in frame {\it a}). It appears that for positive $\mu$
and very large values of $\tan\beta$ the branching ratio in mSUGRA is
unlikely to be at the SM value. Of course, when $m_{1/2}$ (and hence
$m_A$) is very large, the branching ratio tends to the SM value. The gap
is absent for $\mu < 0$. The existence of this gap can be qualitatively
understood by focussing on the chargino contribution (proportional to
the $a_u$ term) and recognizing that models with positive $\mu$ allow
both signs of $A_t$, while models with negative $\mu$ almost always have
$A_t < 0$. As a result, the SUSY contribution may interfere
constructively or destructively with the SM contribution when $\mu > 0$,
while for negative values of $\mu$ the interference is almost always
destructive. Of course, the interference is important only when the SUSY
and SM amplitudes have comparable magnitudes. For the large values of
$\tan\beta$ in this figure, this happens only if $m_A$ is large
(otherwise the SUSY amplitude is much larger than the SM one). In this
case, the coefficients $c_{Q_1}$ and $c_{Q_2}$ in Eq.~(\ref{eq:width})
have the same magnitide.\footnote{This really follows because of the
properties of the MSSM Higgs sector, and so should be true in other
models also.} In units where the SM contribution in the
square parenthesis of Eq.~(\ref{eq:width}) is 1, if we write the
$c_{Q_2}$ contribution as $x = -c_{Q_2}m_{B_s}/2c_{10}m_{\mu}$, the
partial width is determined by $F = x^2 + (1+x)^2$. This has a minimum of
1/2, explaining why the branching fraction does not fall below about
half its SM value. Moreover, as long as $x$ is positive and not small
(as is the case for small values of $m_{1/2}$) $F$ significantly exceeds
the SM value. If $x < 0$, $F$ becomes large only if $|x|$ is very large,
thereby accounting for the gap for positive values of $\mu$. For
negative values of $\mu$, there is no ``positive $x$ branch'', and
hence, no gap. 

\item Although we have not shown this, the scatter plot for $\tan\beta=50$,
$\mu > 0$ is very similar to frame {\it a}), except that the ``gap'' is
somewhat less pronounced. In other words, the difference between the
distribution of dots and crosses in frames {\it b}) and {\it c}) is
absent.

\end{enumerate}

\subsection{Minimal gauge-mediated SUSY breaking model (mGMSB)}

In these models, SUSY breaking is again assumed to occur in a hidden
sector which somehow couples to a set of messenger particles that couple
not only to this hidden sector, but also have SM gauge interactions. The
coupling to the hidden sector induces SUSY breaking in the messenger
sector, which is then conveyed to superpartners of usual particles via
SM gauge interactions.  Messenger particles are assumed to occur in
$n_5$ complete vector representations of $SU(5)$ with quantum numbers of
$SU(2)$ doublets of quarks and leptons. We assume $n_5 \leq 4$ since
otherwise gauge couplings do not remain perturbative up to the scale of
grand unification. The messenger sector mass scale
is characterized by $M$.  The soft SUSY breaking masses for the SUSY
partners of SM particles are thus proportional to the strength of their
corresponding gauge interactions, so that squarks are heavier than
sleptons. Gaugino masses satisfy the usual ``grand
unification'' mass relations, though for very different reasons. Within
the minimal version of this framework, the couplings and masses of the
sparticles in the observable sector are determined (at the messenger
scale $M$) by the parameter set,
\begin{equation}
\Lambda,M,n_5,\tan\beta,sign (\mu ),C_{grav}.
\label{parset}
\end{equation}
$\Lambda$ sets the scale of sparticle masses and is the
most important of these parameters. The
parameter $C_{grav} \geq 1$ and enters only into the
partial width for sparticle decays to the gravitino
and is irrelevant for our analysis. The model predictions for soft-SUSY
breaking parameters at the scale $M$ are evolved to the weak scale using
ISAJET and used to compute $B(B_{q'} \to \ell^+\ell^-)$ as before. 

The novel feature of the GMSB framework is that SUSY breaking can be a
relatively low energy phenomenon if the messenger scale is small. It is
in this case that the gravitino can be an ultra-light LSP. In such a scenario
heavy sparticles cascade decay to the next lightest supersymmetric
particle (NLSP) which then decays to the gravitino and an ordinary
particle. The NLSP may be the lightest neutralino or the
stau.\footnote{For low $\tan\beta$ values not of interest to us, the
sleptons of different generations are essentially degenerate leading to
the so-called co-NLSP scenario\cite{gmsbrep}.} SUSY signatures at the
Tevatron\cite{gmsbrep} are sensitive to the nature of the NLSP. In
our considerations we will focus on models with relatively low messenger
scales for which the collider phenomenology differs the most from mSUGRA.

An important difference between this framework and the mSUGRA model is
that $A$-parameters (at the scale $M$) which are generated only at two
loops are much smaller than scalar and gaugino masses. The weak scale
parameter $A_t$ that is obtained by renormalization group evolution from
the messenger scale is typically smaller than in mSUGRA if the messenger
scale is not large. We thus expect that $a_u$ that sets the scale of the
chargino contributions in (\ref{eq:lageff}) is significantly smaller
than in the mSUGRA framework. We also expect that the gluino
contribution to the loop decay of $B_{q'}$ would also be smaller than in
mSUGRA because there is not enough room to run and induce large mass splitting
between the different $Q_L$s by renormalization group evolution if the
messenger scale is low. Therefore, we typically expect lower values of
$B(B_{q'} \to \ell^+\ell^-)$ relative to corresponding results for mSUGRA
in these scenarios. This branching fraction would, for the same reasons,
increase logarithmically with increasing messenger scale.

Our results for the branching fraction within the GMSB framework are
shown in the $\Lambda-\tan\beta$ plane in Fig.~\ref{fig:gmsbcont}. Here,
we fix $M=3\Lambda$, $n_5=1$ and illustrate the results for both signs
of $\mu$. As before, the dark-shaded region is excluded by the
theoretical considerations on EWSB discussed previously. The bulge on
the left comes from $m_{\ttau_1}^2 < 0$, while in the horizontal strip
at high $\tan\beta$ that extends to large $\Lambda$ values, $m_A^2 < 0$.
In the slant-shaded region $m_{\ttau_1} < 76$~GeV, while in the slivers
covered by triangles along the boundary of the horizontal dark-shaded
region, $m_A < 100$~GeV. Finally, values of $\Lambda$ up to $\Lambda
\sim 53$~TeV (covered by open circles) are excluded because $m_{\te_R} <
100$~GeV, and possibly also by constraints on the chargino mass or
$Z^0$ decay properties\cite{z0decays}.  To the left of the dot-dashed
contour labelled $h$, $m_h < 113$~GeV. The branching fraction for the
decay $b \to s\gamma$ is between $(2-5)\times 10^{-4}$ over the entire
plane in frame~{\it a}), while it is in this range to the right of the
corresponding contour in frame {\it b}).  Contours of $B(B_s \to
\mu^+\mu^-)$ and $B(B_d \to \tau^+\tau^-)$ are shown as solid and dashed
lines, respectively. These are labelled by the value of the
corresponding branching fraction. We see that over almost the entire
plane $B(B_s \to \mu^+\mu^-)$ ($B(B_d \to \tau^+\tau^-)$) is smaller
than $10^{-8}$ ($10^{-7}$) confirming our earlier expectation that the
branching fractions will be smaller than in mSUGRA. It would be
difficult to probe these decays at this level at existing facilities. It
is only for the largest values of $\tan\beta$ where we approach the
region where $m_A^2$ dives to very small values that these branching
fractions might be accessible: however, of this region in frame~{\it b})
is ``excluded'' by the experimental value of $B(b \to s\gamma)$. In
contrast, direct SUSY searches should be sensitive to $\Lambda$ values
of 118-145~TeV depending on the integrated luminosity that is
accumulated\cite{gmsbrep,gmsbtev}. Direct searches at the LHC will
easily probe the entire plane\cite{gmsblhc}.

Up to now, we have assumed $n_5=1$. Since scalar (gaugino) masses scale
with $\sqrt{n_5}$ ($n_5$) we would, expect large sensitivity to $n_5$ if
we show the results in terms of $\Lambda$. To avoid a proliferation of
figures, in Fig.~\ref{fig:gmsbn5} we show the branching fractions versus
$m_{\tg}$ for $n_5=1-4$. We have fixed $M=3\Lambda$ and $\tan\beta=40$
and illustrated the results for both signs of $\mu$. The curves are
terminated at values of $m_{\tg}$ that violate any of the constraints
discussed previously. For very large values of $m_{\tg}$ the branching
fractions approach the SM values because $m_A$ tends to be large. For
smaller values of $m_{\tg}$ the branching fractions (for a given value
of $m_{\tg}$) do depend on $n_5$. Nevertheless, it seems that {\it the
range over which these vary is insensitive to the choice of $n_5$}, so that
our general conclusions drawn from parameter scan for $n_5=1$ should,
broadly speaking, remain unaltered.

\subsection{Minimal Anomaly-Mediated SUSY Breaking Model (AMSB)}

Within the supergravity framework, 
sparticle masses always receive loop contributions
originating in the super-Weyl anomaly when SUSY
is broken. These loop contributions are generally much smaller than
tree level masses. There are, however, classes of models where these loop
contributions may dominate\cite{amsb}. These include
models where there are no SM gauge singlet superfields that can
acquire a Planck scale $vev$ and the usual supergravity contribution to
gaugino masses is suppressed by an additional factor $\frac{M_{SUSY}}
{M_P}$ relative to $m_{\frac{3}{2}} = M_{SUSY}^2/M_P$, or  higher
dimensional models where the coupling between the observable and hidden
sectors is strongly suppressed.

The anomaly-mediated SUSY breaking contributions to gaugino masses are
proportional to the corresponding gauge group $\beta$-function, and so
are non-universal.  Likewise, scalar masses and trilinear terms are
given in terms of gauge group and Yukawa interaction beta functions. As
a result sparticles with the same gauge quantum numbers have a common
mass so that flavour changing effects from this sector are naturally
small.  Slepton squared masses, however, turn out to be negative
(tachyonic). A ``minimal'' fix, that does not upset the resolution of
the flavour problem is to assume an additional contribution
$m_0^2$ for all scalars.  Since the magnitude of $\mu$ is fixed by the
observed value of $M_Z$, the parameter space of the model then consists
of
\begin{equation}
m_0,\ m_{3/2},\ \tan\beta\ {\rm and}\ sign(\mu ) .
\end{equation}
Weak scale SUSY parameters can now be computed via renormalization group
evolution, and $B(B_{q'} \to \ell^+\ell^-)$ can again be calculated
using (\ref{eq:width}). 

For small values of $m_0$, the scale of sparticle masses is set by
$m_{3/2}$ which cannot be much smaller than about 35~TeV. In
Fig.~\ref{fig:amsbtb} we show the branching ratio for $B_s
\to \mu^+\mu^-$ and $B_d \to \tau^+\tau^-$ decays versus $\tan\beta$,
with $m_{3/2}=40$~TeV and {\it a})~$m_0=400$~GeV, and {\it
b})~$m_0=1$~TeV. The solid (dashed) curves correspond to positive
(negative) values of $\mu$. As before, the squares mark the
experimentally allowed upper limit on $\tan\beta$. Also, shown by the
right hand scale in the
figure, is the branching fraction for the decay $b \to s\gamma$. 

We see from Fig.~\ref{fig:amsbtb} that for positive values of $\mu$ the
$B(B_{q'} \to \ell^+\ell^-)$ is close to its SM value, and relatively
insensitive to the value of $\tan\beta$. For negative values of $\mu$,
these branching fractions can indeed become large, again when $m_A$
dives to low values. However, for the range of $\tan\beta$ where $B(b
\to s\gamma) < 5\times 10^{-4}$, the branching fraction for $B_s \to
\mu^+\mu^-$ decay will be difficult to detect in Tevatron
experiments. We have also examined the dependence of these branching
fractions on $m_{3/2}$. For positive values of $\mu$, and
$\tan\beta=40-60$, they are once again always close to the corresponding
SM value, independent of $m_{3/2}$. For negative values of $\mu$, $B(B_s
\to \mu^+\mu^-)$ increases rapidly with $m_{3/2}$, and even for
$\tan\beta=40$ and $m_0=1$~TeV can be as large as $10^{-5}$; however,
for the range of $m_{3/2}$ where $B(b \to s\gamma)$ is not too large,
the branching fraction can be as large as several times $10^{-8}$ and
may be accessible at the Tevatron, but would require considerable integrated
luminosity. 

In Fig.~\ref{fig:amsbcont}, we show contours for $B(B_{q'}
\to \ell^+\ell^-)$ in the $m_0-m_{3/2}$ plane, where we have fixed
$\tan\beta=42$. We show results only for $\mu < 0$ since for positive
values of $\mu$ the branching fraction appears to be close to the SM one
even for very large values of $\tan\beta$. We remind the reader that in
the AMSB framework negative $\mu$ is also favoured by the result of the
E821 experiment\cite{g-2exp}. The dark-shaded region is excluded by
theoretical constraints
and in the open circle region $m_{\tw_1} < 100$~GeV. Along the line of
squares and triangles running diagonally across the figure, $m_h$ or
$m_A$, respectively fall below their experimental bounds. Finally,
except in the narrow region covered with dots that follows the boundary
of the upper shaded region where $B(b \to s\gamma)$ is too large, the
branching fraction for the decay $b\to s\gamma$ is in its ``allowed
range'' of $(2-5)\times 10^{-4}$. The three solid contours correspond to
$B(B_s \to \mu^+\mu^-) = 10^{-8}, 5\times 10^{-8}$ and 10$^{-7}$, while
the dashed contours correspond to $B(B_d \to \tau^+\tau^-) = 10^{-7},
10^{-6}$ and 10$^{-5}$. We see that the branching fractions within the
AMSB framework is small over most of the parameter plane. The decay $B_s
\to \mu^+\mu^-$ may be probed in Tevatron experiments only over the
limited range where $m_A$ is diving to relatively low values. With a
large integrated luminosity, there is a limited parameter range
(consistent with other collider constraints) where this decay may be the
harbinger of new physics at the Tevatron.  With just 10~fb$^{-1}$ of
integrated luminosity direct searches for sparticles at the LHC should
be sensitive\cite{amsblhc} to $m_{3/2} \sim 70-90$~TeV, depending on the
value of $m_0$.

For larger values of $\tan\beta$ and $\mu < 0$, the situation is
qualitatively similar except that the region excluded by the theoretical
constraints expands leaving a yet narrower wedge of ``allowed
parameters''. 

\subsection{$SU(5)$ models with non-universal gaugino masses}

Since supergravity is not a renormalizable theory, there is no reason to
suppose that the gauge kinetic function $f_{ab} =\delta_{ab}$.  Indeed,
if the gauge kinetic function develops a SUSY breaking $vev$ that also breaks
the GUT symmetry, non-universal gaugino masses result.
Expanding the gauge kinetic function as $f_{ab} = \delta_{ab} +
\hat{\Phi}_{ab}/\mpl + \ldots$, where the fields $\hat{\Phi}_{ab}$
transform as left handed chiral superfields under supersymmetry
transformations, and as the symmetric product of two adjoints under
gauge transformations, we parametrize the lowest order contribution to
gaugino masses by,
\begin{equation}
{\cal L}\supset \int d^2\theta {\hat{W}}^a{\hat{W}}^b
{\hat{\Phi}_{ab}\over M_{\rm Planck}} + h.c.
\supset  {\langle F_{\Phi} \rangle_{ab}\over M_{\rm Planck}}
\lambda^a\lambda^b\, +\ldots ,
\end{equation}
where ${\hat{W}}^a$ is the superfield that
contains the gaugino field $\lambda^a$,
and $F_{\Phi}$ is the auxillary field component of $\hat{\Phi}$ that
acquires a SUSY breaking $vev$.
In principle, the chiral superfield $\hat{\Phi}$
which communicates supersymmetry
breaking to the gaugino fields can lie in any representation 
contained in the
symmetric product of two adjoints, and so can lead to gaugino mass terms
that break the underlying gauge symmetry. We require, of course, that SM
gauge symmetry is preserved.

In the context of $SU(5)$ grand unification, $F_{\Phi}$ would most
generally be
a superposition of irreducible representations which appears in the symmetric
product of two ${\bf 24}$s,
\begin{equation}
({\bf 24}{\bf \times}
 {\bf 24})_{\rm symmetric}={\bf 1}\oplus {\bf 24} \oplus {\bf 75}
 \oplus {\bf 200}\,,
\label{irrreps}
\end{equation}
where only $\bf 1$ yields universal gaugino masses as in the mSUGRA
model.  The relations amongst the various GUT scale gaugino masses have
been worked out {\it e.g.} in Ref. \cite{anderson}, and are listed in
Table~\ref{masses} along with the approximate masses after RGE evolution
to $Q\sim M_Z$.  Motivated by the measured values of the gauge couplings
at LEP, we assume that the $vev$ of the SUSY-preserving scalar component
of ${\hat{\Phi}}$ is neglible.  Each of these three non-singlet models
is as predictive as the canonical singlet case, and all are
compatible with the unification of gauge couplings.  Although
superpositions are possible, for definiteness we only consider the
predicitive subset of scenarios where $F_{\Phi}$ transforms as one of
the irreducible representations of $SU(5)$.  The model parameters may
then be chosen to be,
\begin{equation}
m_0,\ M_3^0,\ A_0,\ \tan\beta\ {\rm and}\ sign(\mu ),
\end{equation}
where $M_i^0$ is the $SU(i)$ gaugino mass at scale $Q=M_{GUT}$: $M_2^0$
and $M_1^0$ can then be calculated in terms of $M_3^0$ using to
Table \ref{masses}. The sparticle masses and mixing angles can then
be computed using the {\em non-universal SUGRA} option in ISAJET. An
illustrative example of the spectrum may be found in Ref.\cite{quintana}
where this framework is reviewed.
We see from Table~\ref{masses} that the pattern of gaugino masses at the
weak scale differs quite significantly from mSUGRA expectation. This
suggests that the relative contribution of the chargino and
gluino-mediated contributions to $B_s \to \mu^+\mu^-$ decays, and hence
the branching fraction may differ significantly from their values in the
mSUGRA framework. 

To illustrate this, we show $B(B_s\to\mu^+\mu^-)$ as a function of the
input GUT scale gluino mass parameter in Fig.~\ref{fig:nunivm30} for the
various models in Table~\ref{masses}. We fix $m_0=400$~GeV, $A_0=0$,
$\tan\beta=40$ and show our results for both signs of $\mu$. The dotted
line labelled 1 DR shows the branching fraction within the mSUGRA
framework where $\Phat$ transforms as a singlet of $SU(5)$, while the
solid and dashed lines show the result for the cases where $\Phat$
transforms as a {\bf 24} or {\bf 200} dimensional representation,
respectively. The {\bf 75} case is theoretically excluded for our choice
of parameters. Indeed, we see that the branching fraction is sensitive
to the underlying pattern of non-universality.\footnote{Values of
$M_3^0$ smaller than 200~GeV (300~GeV) are ``excluded'' for the {\bf
200} ({\bf 24}) model since these yield $m_h < 113$~GeV.}  Especially
striking is the dip at $M_3^0 \simeq 200$~GeV in frame {\it b}). We have
checked that this is due to an almost complete cancellation between the
chargino and gluino loops, which as far as we can ascertain is
completely accidental. Individually, either the chargino or gluino loop
contribution would be much larger than the SM one ($m_A$ is only
175~GeV), but the total SUSY amplitude is comparable to the SM one and
the two interfere destructively to yield a branching fraction which is
close to its minimum value (recall our discussion at the end of
Sec.~\ref{msugrares}). This case highlights the importance of including
the gluino loop, without which the branching fraction would be almost an
order of magnitude bigger. We should also mention that contributions
from neutralino loops that have been neglected in our analysis may now be
significant, in which case $B(B_d \to \tau^+\tau^-)$ would not have the
canonical ratio with $B(B_s \to \mu^+\mu^-)$ \cite{bobeth2}. 

To see how large $B(B_s \to \mu^+\mu^-)$ could be in these scenarios, we
scanned over the parameter space of these models. We allow $m_0$ to vary
between 100~GeV and $\sqrt{50}M_3^0$ and $-3m_0 < A_0 < 3A_0$. In
Fig.~\ref{fig:nunivscanp}, where we take $\tan\beta=55$, we show results
for $\mu > 0$ for the three cases of non-universality discussed
above. The corresponding result for the mSUGRA case was shown in
Fig.~\ref{fig:sugscan}{\it a}. Analogous results for $\mu < 0$ and
$\tan\beta=45$ are shown in Fig.~\ref{fig:nunivscanm} with the
corresponding result for the mSUGRA case now in
Fig.~\ref{fig:sugscan}{\it b}. Again, points that satisfy other
experimental constraints\footnote{In the case of the {\bf 75} and {\bf
200} models the mass gap between the chargino and the LSP is small, and
LEP constraints on $m_{\tw_1}$ cited here may be too stringent; see,
however, the discussion in Ref.\cite{nonuniv}.}  but where $B(b \to
s\gamma)$ falls within (outside) the range $(2-5)\times 10^{-4}$ are
denoted by a cross (dot). The following features of these figures are
worth noting:
\begin{enumerate}
\item For the {\bf 24} model, the branching fraction is always smaller
than about $2\times 10^{-8}$ at least for the $\tan\beta$ values shown,
and would be difficult to probe at the Tevatron. However, in this
scenario that Tevatron experiments should see signals in the $\eslt$,
$jets + 1\ell +\eslt$, and possibly, $jets + Z^0\to \ell^+\ell^- +\eslt$
channels if $M_3^0 \alt 175$~GeV and $m_0 \alt 400-500$~GeV
\cite{nonuniv}. 
For $\mu < 0$ \cite{ourg-2} favored by the measured
value of the muon anomalous magnetic moment, the branching fraction
tends to be within a factor $\sim 2$ of its SM value.

\item The branching fraction for $B_s \to \mu^+\mu^-$ decays can be
large for the {\bf 75} model, regardless of the sign of $\mu$. For
positive values of $\mu$ favoured by the E821 experiment\cite{g-2exp},  
$B(B_s\to\mu^+\mu^-)$ can be as large as $10^{-6}$, which is just a
little over a factor of 2 from its current upper limit. The gap in 
Fig.~\ref{fig:nunivscanp}{\it b} is presumably for the same reason as
in the mSUGRA case. It is, however, more pronounced because small values 
of $|A_0|$ are not allowed. For the {\bf 75} model, direct SUSY searches 
at the Tevatron can give observable signals only in the $\eslt$ channel,
and that too only if $M_3^0$ and $m_0$ are rather small. In this case, 
$m_{\tw_1}, m_{\tz_2}$ and $m_{\tz_1}$ are all rather close in mass, so
that decays of $\tw_1$ and $\tz_2$ typically lead to very small visible
energy\cite{nonuniv}. 

\item For the {\bf 200} model, $B(B_s\to\mu^+\mu^-)$ can exceed
$10^{-7}$, 
even for very large values of $M_3^0$ for both signs of $\mu$ and
parameter values consistent with the measured value of $B(b \to
s\gamma)$. As in the {\bf 75} case, direct SUSY signals at the Tevatron,
for the most part, will be restricted to the $\eslt$ channel\cite{nonuniv}.

\end{enumerate}

\section{Summary} \label{summary}

If $\tan\beta$ is large, substantial flavour violating couplings of
neutral MSSM Higgs bosons to down type quarks are induced at the 1-loop
level via diagrams with squarks and charginos or squarks and
gluinos in the loop. These induced interactions, in turn, lead to new Higgs
boson mediated SUSY contributions to the amplitude for the decays
$B_s\to \mu^+\mu^-$ and $B_d\to \tau^+\tau^-$. Depending on model
parameters, the branching fraction for the decays may be several
orders of magnitude larger than its corresponding SM
expectation. Moreover, while the SUSY contribution to the decay 
decouples when $m_A$ becomes large (so that the Higgs sector of the MSSM
reduces to that of the SM with a light Higgs boson), it does not
decouple for heavy sparticles. 

The CDF experiment has already established that $B(B_s \to\mu^+\mu^-)
\leq 2.6\times 10^{-6}$, and with the data sample of $\sim 2$~fb$^{-1}$
that is expected to be collected after 2-3 years of Main Injector
operation, will be sensitive to branching fractions smaller than about
$10^{-7}$. The sensitivity will be even greater as Tevatron experiments
approach their goal of $\sim 15$~fb$^{-1}$.  This then opens up the
possibility of discovering new physics at the Tevatron even if
sparticles are heavy and their signals from direct production below the
level of observability. 

These considerations led us to examine predictions for $B(B_s
\to\mu^+\mu^-)$. This is, however, not possible within the context of
the generic MSSM since SUSY amplitudes are sensitive to the {\it a
priori} unknown flavour structure of squark mass matrices and trilinear
couplings. In other words general predictions (as a function of
sparticle masses) are not possible, and one has to resort to specific
models. In this paper, we have examined how this branching ratio within
the framework of several popular models of SUSY breaking: the mSUGRA
model, the minimal gauge mediated SUSY breaking model, and the
anomaly-mediated SUSY breaking model.\footnote{We did not separately
examine the gaugino-mediated SUSY breaking
framework\cite{inomediation}. This is a model based on extra dimensions
where usual matter and SUSY breaking fields reside on different
spatially separated branes, while gauge fields live in the bulk. As a
result, gauginos which directly ``feel'' the SUSY breaking develop
masses at the compactification scale $M_c$, while matter scalars do
not. Renormalization effects then induce scalar masses and
$A$-parameters at $Q = M_{GUT} < M_c$. Although some $GUT$ scale
non-universality is induced because of Yukawa couplings, we expect that
results in this framework would be qualitatively similar to those in
mSUGRA where $m_0, -A_0 \alt 0.5 m_{1/2}$. For a discussion of the
collider phenomenology of these models, see Ref.\cite{inomsb}.}  We also
examined the range of this branching ratio in supergravity $SU(5)$
models with non-universal gaugino masses.  Specifically, we studied how
the $B(B_s \to\mu^+\mu^-)$ varies with parameters, and delineated
regions of parameter space where it might be observable at the
Tevatron. Without making any representation about the sensitivity of
BELLE and BABAR, we have also examined the branching fraction for the
related decay $B_d \to \tau^+\tau^-$. Since the coupling of the Higgs to
SM fermions is proportional to the fermion mass, the partial width for
this decay is enhanced (relative to that for $B_s\to\mu^+\mu^-$) by a
factor $(m_{\tau}/m_{\mu})^2$, but reduced by a factor
$(|V_{td}|/|V_{ts}|)^2$ that originates in the flavour violating Higgs
boson vertex. Our rule of thumb is that these experiments will be
competitive with Tevatron experiments if their sensitivity to $B(B_d \to
\tau^+\tau^-)$ is no more than an order of magnitude worse than the
Tevatron sensitivity to $B(B_s \to \mu^+\mu^-)$.

Our main results are summarized in
Fig.~\ref{fig:sugcont}--\ref{fig:sugscan} for mSUGRA, in
Figs.~\ref{fig:gmsbcont} and Fig.~\ref{fig:gmsbn5} for the GMSB model,
in Fig.~\ref{fig:amsbcont} for the AMSB model, and in
Fig.~\ref{fig:nunivscanp} and Fig.~\ref{fig:nunivscanm} for $SU(5)$
models with non-universal GUT scale gaugino masses.  We will not
elaborate on the details here. Generally speaking, in the GMSB and AMSB
frameworks, the branching fraction for this decay is significantly
smaller than in the mSUGRA model. Within the GMSB and AMSB scenarios,
the reach of Tevatron experiments via this decay will be relatively
limited. A branching fraction close to $10^{-7}$ is possible only for
regions of parameters where $m_A$ tends to be small due to accidental
cancellations. In supergravity models with non-universal gaugino masses,
the size of the signal is sensitive to the gaugino mass pattern. Indeed,
in the {\bf 75} and {\bf 200} models the signal can be very large for
values of parameters consistent with other constraints.  In all these
models there are varying regions of parameters that are not in conflict
with observations, where $B_s \to \mu^+\mu^-$ should be observable at
the Tevatron, and where there are no direct sparticle or Higgs boson
signals.

The decays $B_s \to \mu^+\mu^-$ and $B_d \to \tau^+\tau^-$ may also be
incisive probes of $SO(10)$ SUSY models with Yukawa coupling
unification since these require $\tan\beta$ to be large. These models
have recently received considerable attention\cite{soten}, especially in
light of the interpretation of the atmospheric neutrino
data\cite{superk} as neutrino oscillations originating in a non-trivial
flavour structure of the neutrino mass matrix. Since Yukawa coupling
unification is possible only in very special regions of parameter space that
require extensive scanning to find, we  defer the analysis of $B(B_{q'} \to
\ell^+\ell^-)$ within this framework to a dedicated study of the
phenomenology of these models that is currently in progress.

Before concluding, we again emphasize that the branching fractions that
we have examined are sensitive to details of the model in that small
deviations in some of the assumptions (about sfermion flavour structure
at the high scale) that define the framework could result in
considerable changes in the answer. Our point here is that we should use
various ``model predictions'' only for guidance, or as benchmarks, but
keep open the possibility that the real world may be somewhat
different. By the same token, accurate determination of flavour
violating processes serves as a sensitive probe of flavour structure at
very high scales. The observation of processes such as $B_s \to
\mu^+\mu^-$ at the Tevatron or $B_d \to \tau^+\tau^-$ at $B$-factories
is important not only because it would herald new physics, but also
because they may serve as probes of flavour physics at scales not
directly accessible to experiment.

\section*{Acknowledgments}
It is a pleasure to thank C.~Kolda for a conversation 
about Ref.\cite{babu} that got us started on this study. 
We thank H.~Dreiner, U.~Nierste and especially A.~Dedes for
correspondence and detailed comparisons of their results against
ours. We are grateful to H.~Baer and J.~Ferrandis for their help in
improvements in the calculation of $m_A$ using ISAJET. But for their
intervention, this paper may have been released considerably earlier.
We thank E.~Polycarpo for valuable information about the sensitivity of
LHC experiments to $B_s \to \mu^+\mu^-$ decays, and for providing us
with Ref.\cite{lhc}. 
This research was supported in part by the U.S. Department of Energy
under contracts number DE-FG02-97ER41022 and by Funda\c{c}\~ao de Amparo
\`a Pesquisa do Estado de S\~ao Paulo (FAPESP).


%
\begin{table}[htb]
\begin{center}
\begin{small}
\begin{tabular}{c|ccc|ccc}
\ & \multicolumn{3}{c|} {$\mgut$} & \multicolumn{3}{c}{$\mz$} \cr
$F_{\Phi}$
& $M_3$ & $M_2$ & $M_1$
& $M_3$ & $M_2$ & $M_1$ \cr
\hline
${\bf 1}$   & $1$ &$\;\; 1$  &$\;\;1$   & $\sim \;6$ & $\sim \;\;2$ &
$\sim \;\;1$ \cr
${\bf 24}$  & $2$ &$-3$      & $-1$  & $\sim 12$ & $\sim -6$ &
$\sim -1$ \cr
 ${\bf 75}$  & $1$ & $\;\;3$  &$-5$      & $\sim \;6$ & $\sim \;\;6$ &
$\sim -5$ \cr
${\bf 200}$ & $1$ & $\;\; 2$ & $\;10$   & $\sim \;6$ & $\sim \;\;4$ &
$\sim \;10$ \cr
\end{tabular}
\end{small}
\smallskip
\caption{Relative gaugino masses at $\mgut$ and $\mz$
in the four possible irreducible representations that  $F_{\Phi}$ could
transform as.}
\label{masses}
\end{center}
\end{table}
%

%
\iftightenlines\else\newpage\fi
\iftightenlines\global\firstfigfalse\fi
\def\dofig#1#2{\epsfxsize=#1\centerline{\epsfbox{#2}}}
\def\fig#1#2{\epsfxsize=#1{\epsfbox{#2}}}

%
\begin{figure}
\dofig{12cm}{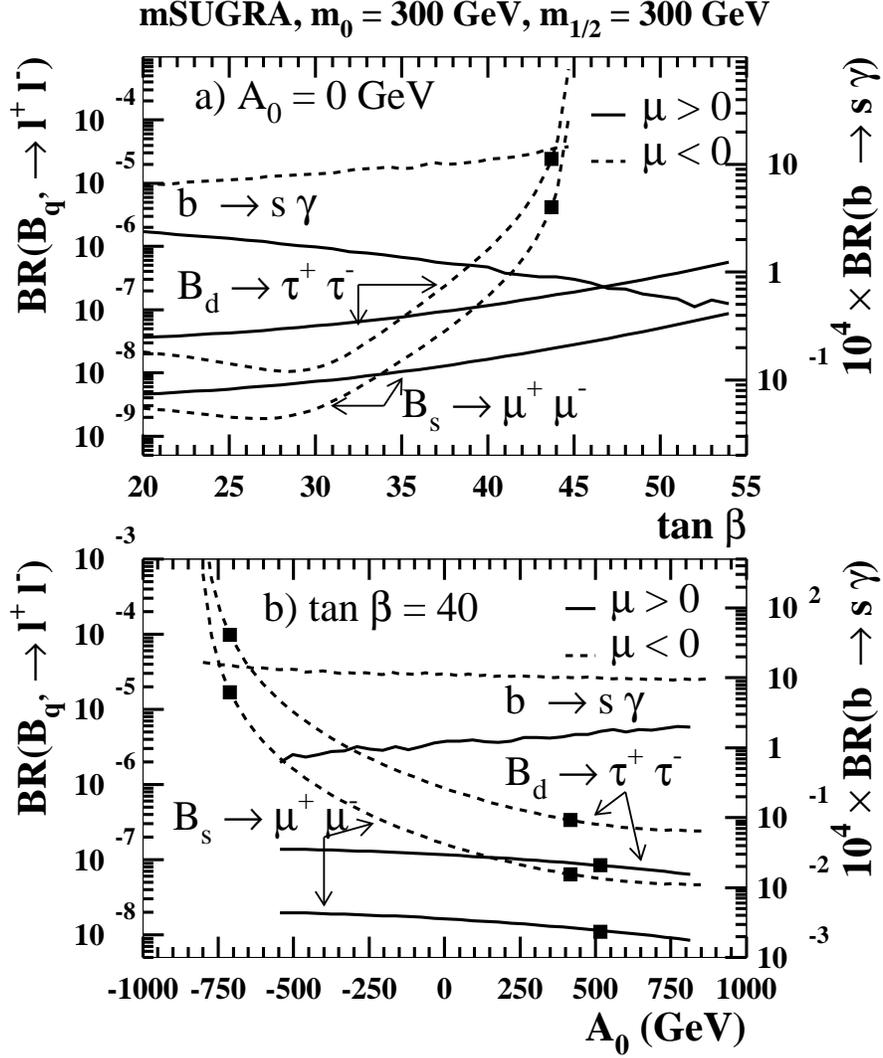}
\caption[]{Branching fractions for the decays $B_s \to \mu^+\mu^-$ and 
$B_d\to \tau^+\tau^-$ in the mSUGRA model with $m_0=300$~GeV and
$m_{1/2}= 300$~GeV, for $\mu>0$ (solid) and $\mu < 0$
(dashed). In frame {\it a}) we show the branching fractions versus
$\tan\beta$  for $A_0=0$, while in frame {\it b}) we show these versus
$A_0$ for $\tan\beta=40$. Also shown is the branching fraction for the 
decay $b \to s\gamma$ which is to be read off using the scale on the
right. The squares mark the limits of the experimentally allowed
regions as discussed in the text. }
\label{fig:sugtanbA}
\end{figure}

\newpage

\begin{figure}
\noindent
\dofig{12cm}{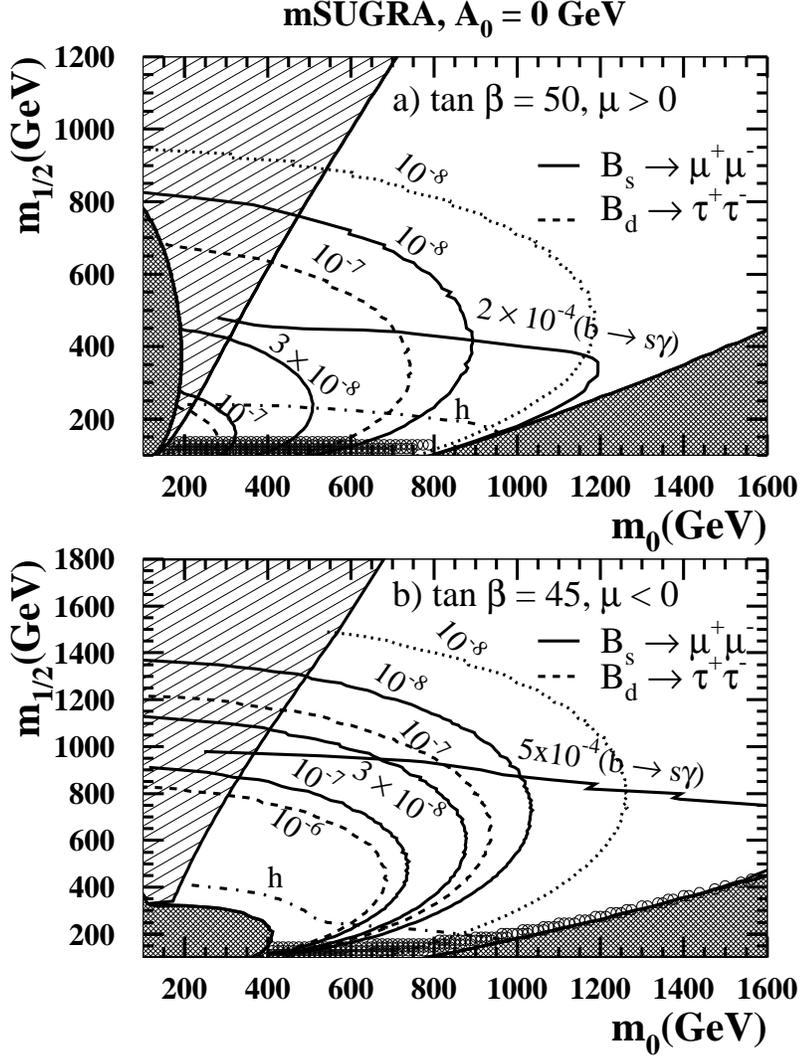}
\caption[]{Contours of constant branching fraction for the decays $B_s
\to\mu^+\mu^-$ (solid) and $B_d\to \tau^+\tau^-$ (dashed) marked on the
curves. The results are shown in the $m_0-m_{1/2}$ plane of the mSUGRA
model with $A_0=0$, for {\it a}) $\tan\beta=50, \mu> 0$, and {\it
b})~$\tan\beta=45, \mu < 0$. The dotted contour labelled $10^{-8}$ shows
the result for $B(B_s \to\mu^+\mu^-)$ if the gluino loop contribution
discussed in the text is set to zero. The dark-shaded region is excluded
by theoretical contraints on EWSB. In the slant-hatched region, $\tz_1$
is not the LSP. The region covered by open circles is excluded by
experimental constraints on sparticle masses or on $m_A$, as discussed
in the text. Below the dot-dashed curve labelled $h$, $m_h< 113$~GeV. In
frame {\it a}) below the solid curve labelled $b\to s\gamma$, $B(b\to
s\gamma) < 2\times 10^{-4}$. Below the corresponding curve in frame {\it
b}), $B(b\to s\gamma) > 5\times 10^{-4}$.  }
\label{fig:sugcont}
\end{figure}

\newpage

\begin{figure}
\dofig{12cm}{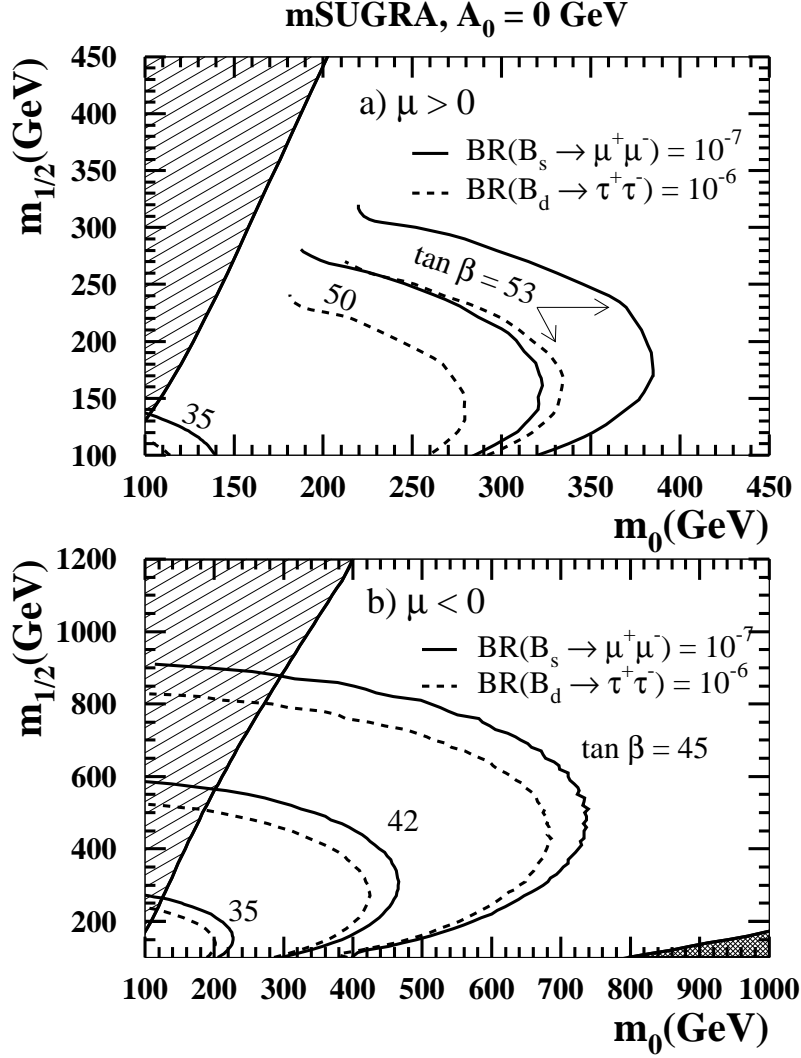}
\caption[]{Contours in the $m_0-m_{1/2}$ plane of the mSUGRA model with
$A_0=0$ where the branching fraction for $B_s\to \mu^+\mu^-$ ($B_d
\to\tau^+\tau^-$) is $10^{-7}$ (10$^{-6}$) for several values of
$\tan\beta$. Shaded regions are as in the previous figure.  In frame
{\it a}) we take $\mu > 0$, while in frame {\it b}) we take $\mu < 0$.}
\label{fig:sugtbcont}
\end{figure}
\newpage

\begin{figure}
\dofig{12cm}{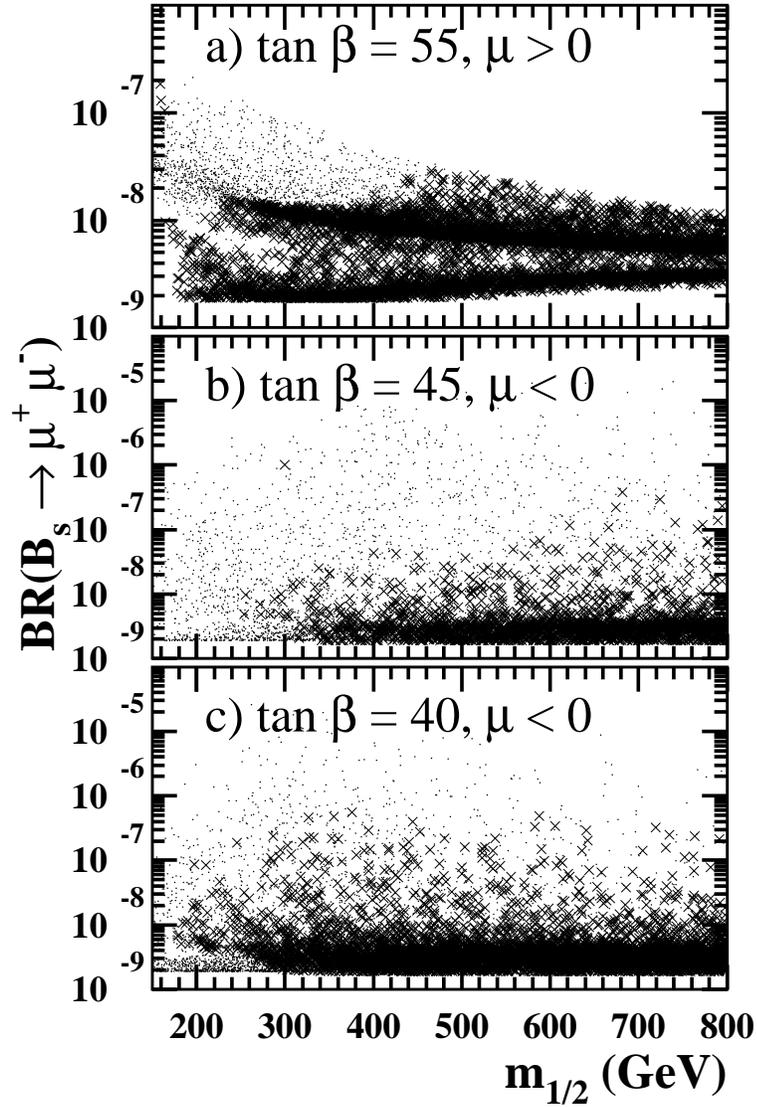}
\caption[]{The branching fraction for the decay $B_s \to \mu^+\mu^-$ for
a scan of the mSUGRA parameter space over the range mentioned in the
text. We show results versus $m_{1/2}$ for {\it a})~$\tan\beta=55, \mu >
0$, {\it b})~$\tan\beta=45, \mu < 0$, and {\it c})~$\tan\beta=40, \mu < 0$.
Each cross (dot) denotes a model where $B(b\to s\gamma)$ lies within
(outside) the range $(2-5)\times 10^{-4}$. }
\label{fig:sugscan}
\end{figure}
\newpage

%
\begin{figure}
\dofig{12cm}{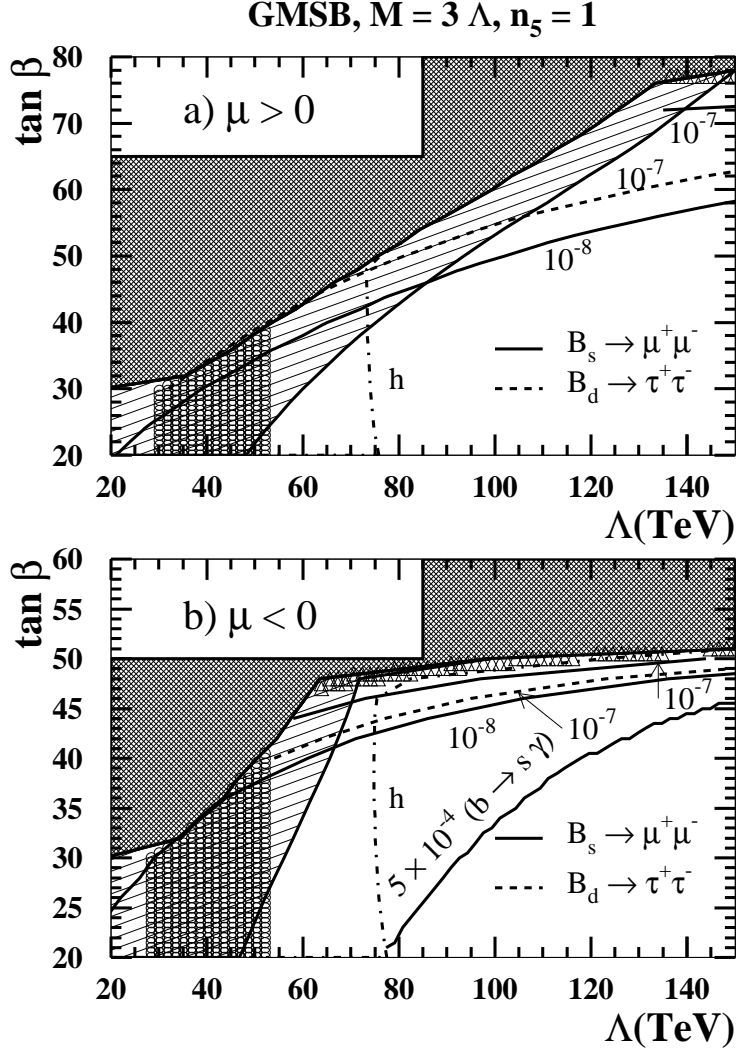}
\caption[] {Contours of constant branching fraction for the decays $B_s
\to\mu^+\mu^-$ (solid) and $B_d\to \tau^+\tau^-$ (dashed) with values
marked on the curves in the $\Lambda-\tan\beta$ plane of the minimal
GMSB model. We take the messenger scale $M=3\Lambda$ and $n_5=1$ in this
figure and show results for {\it a})~$\mu> 0$, and {\it b})~$\mu <
0$. In both frames, the dark-shaded region is excluded by theoretical
considerations discussed in the text. In the slant-hatched region
$m_{\ttau_1} < 76$~GeV, in the region covered by open circles $m_{\te_R}
< 100$~GeV, and in the region covered by triangles, $m_A < 100$~GeV. The
open-circle region extends all the way to the smallest values of
$\Lambda$ (as does the slant-hatched region) but has been terminated at
$\Lambda=30$~TeV for clarity.  To the right of the contour labelled $b
\to s\gamma$ in frame {\it b}), $B(b\to s\gamma)$ is in the range
$(2-5)\times 10^{-4}$. It is in this range all over the plane in frame
{\it a}). }
\label{fig:gmsbcont}
\end{figure}
\newpage

\begin{figure}
\dofig{12cm}{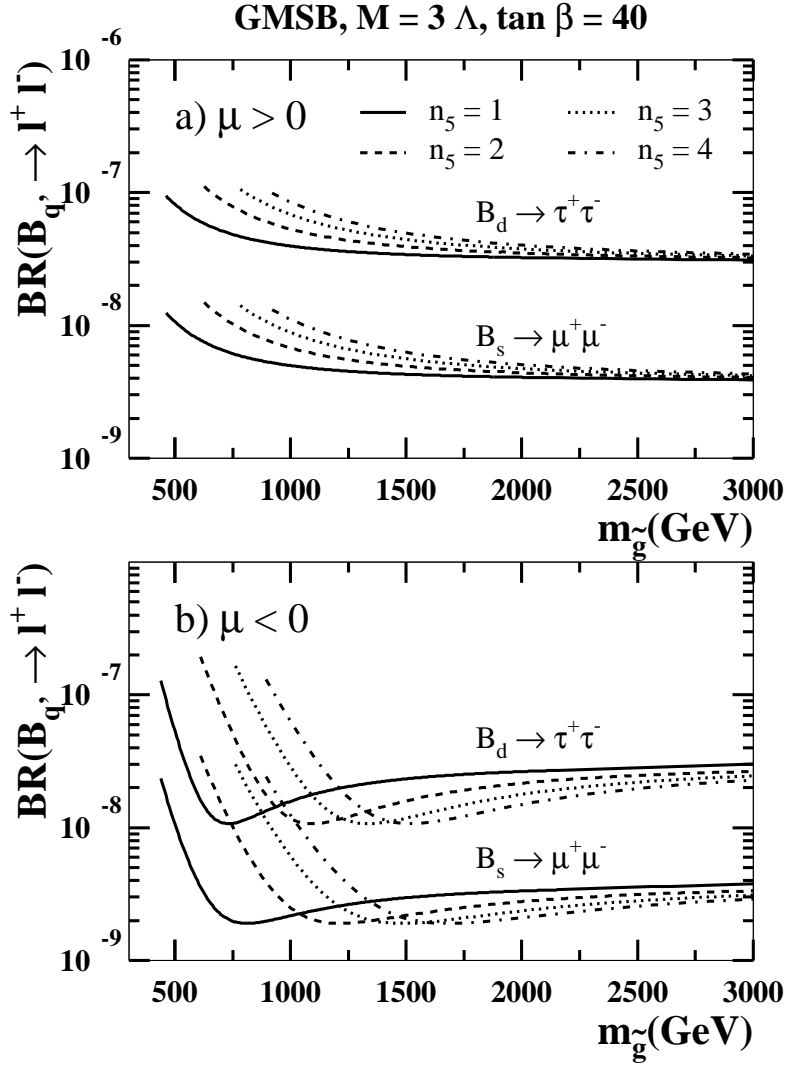}
\caption[]{The dependence of the branching fractions for the decays
$B_s\to\mu^+\mu^-$ and $B_d\to \tau^+\tau^-$ on the parameter $n_5$. We
plot the branching fractions versus a sparticle mass (we choose
$m_{\tg}$) rather than the theoretical parameter $\Lambda$ for reasons
discussed in the text. We show our results for {\it a})~$\mu> 0$, and
{\it b})~$\mu < 0$. Other parameters are as shown on the figure.}
\label{fig:gmsbn5}
\end{figure}
\newpage

%
\begin{figure}
\dofig{12cm}{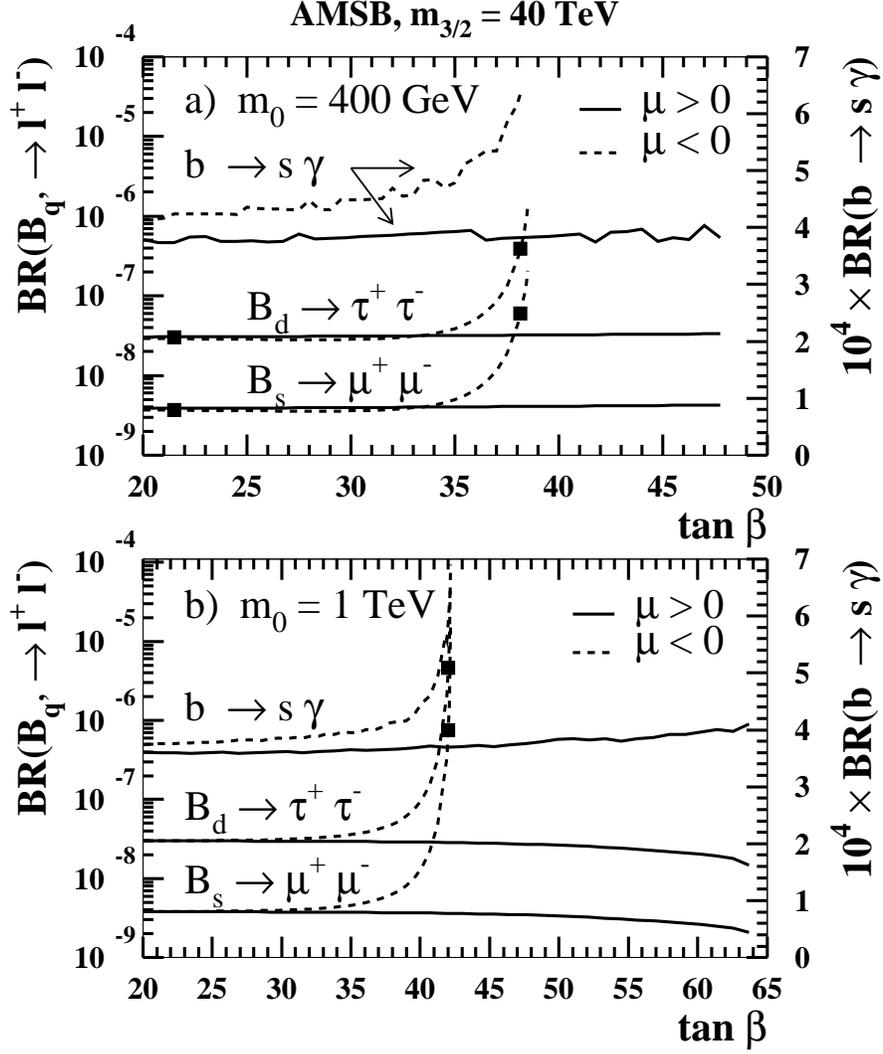}
\caption[]{Branching fractions for the   decays
$B_s\to\mu^+\mu^-$ and $B_d\to \tau^+\tau^-$ within the framework of the
minimal anomaly-mediated SUSY breaking model. We show the branching
fractions versus $\tan\beta$ for {\it a})~$m_0=400$~GeV and {\it
b})~$m_0=1$~TeV, and other parameters as labelled on the figure. Also
shown is the branching fraction for the decay $b\to s\gamma$ which
should be read using the scale on the right. }
\label{fig:amsbtb}
\end{figure}
\newpage

\begin{figure}
\dofig{15cm}{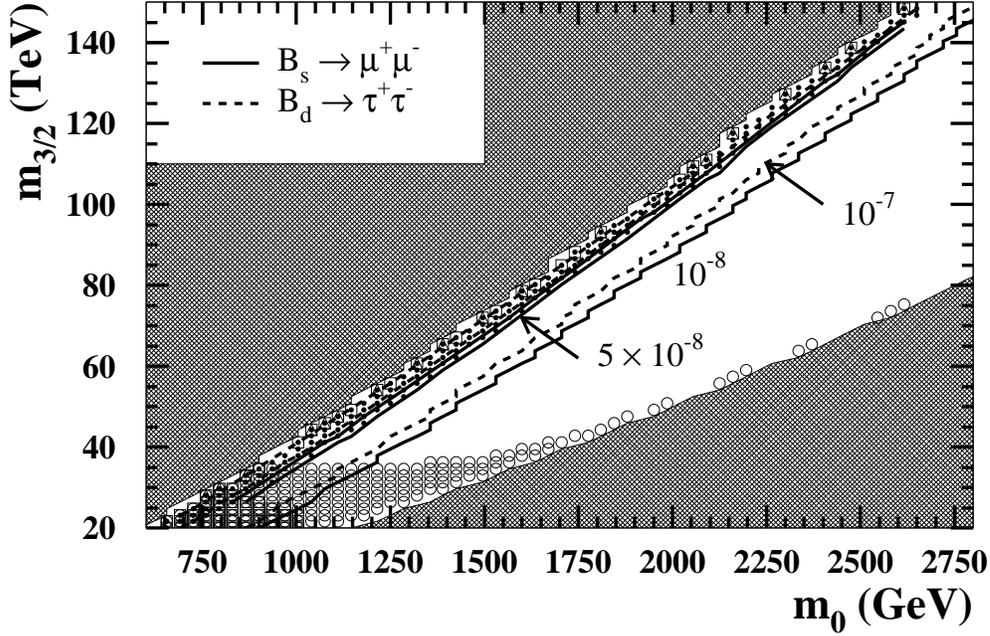}
\caption[]{Contours of branching fraction for the decays
$B_s\to\mu^+\mu^-$ (solid) and  $B_d\to \tau^+\tau^-$ (dashed) in
the $m_0-m_{3/2}$ plane of the  mAMSB model. 
The values of the branching fractions are mentioned in the text.
The dark-shaded region is excluded
by theoretical considerations. Along the squares and triangles running
along the boundary of the upper dark-shaded region, $m_h$ and  $m_A$
respectively fall below their experimental bound. In the region covered
by open circles, $m_{\tw_1}< 100$~GeV. Finally, in the region covered by
dots close to where the squares and triangles are, $B(b\to s\gamma)>
5\times 10^{-4}$, while over the unshaded region it is in the range
$(2-5)\times 10^{-4}$.  }
\label{fig:amsbcont}
\end{figure}
\newpage


\begin{figure}
\noindent
\dofig{12cm}{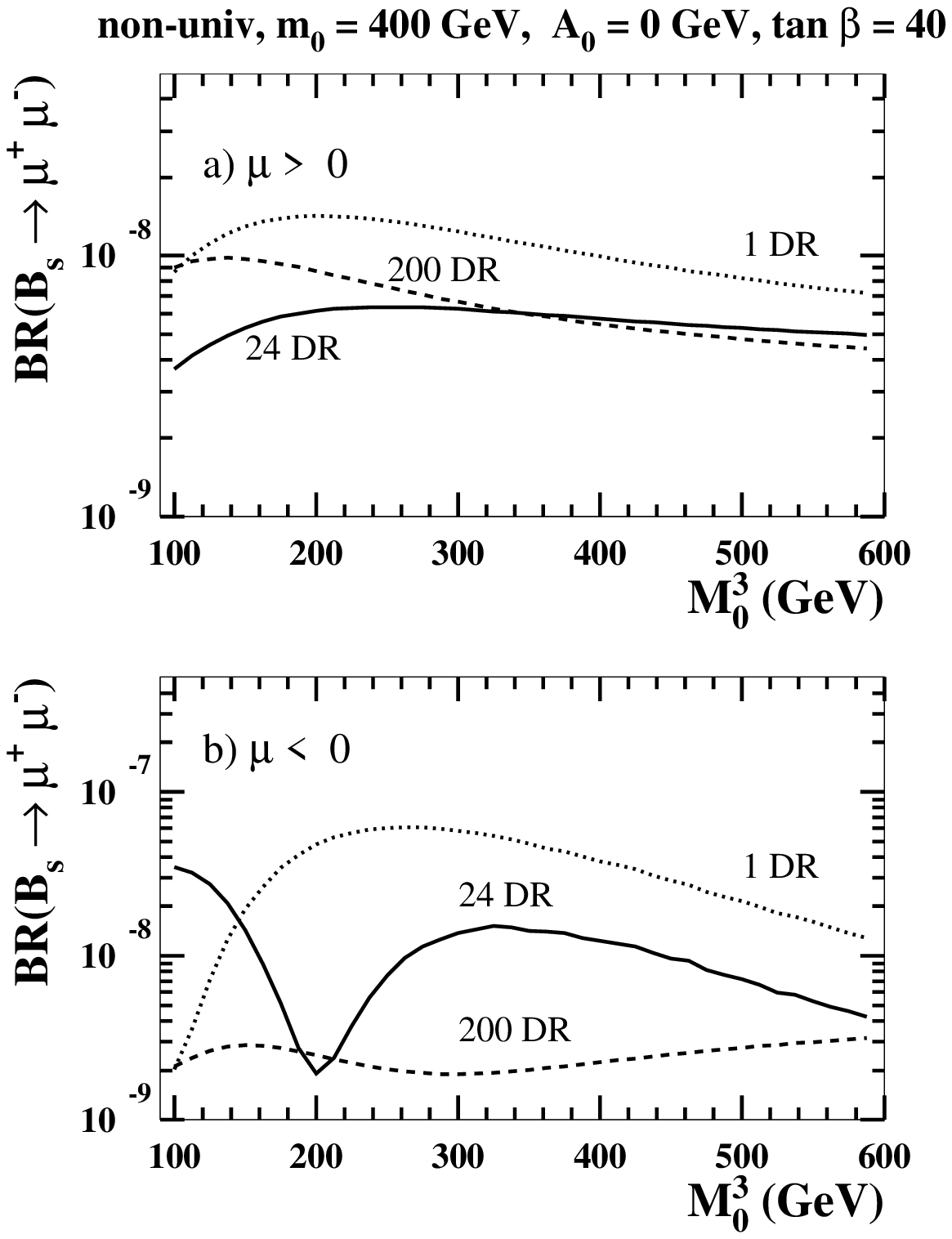}
\caption[]{The branching fraction for the decay $B_s\to\mu^+\mu^-$
versus the GUT scale gluino mass parameter $M_3^0$ in $SU(5)$
supergravity models with non-universal gaugino masses. As discussed in
the text, the models are characterized by the gauge transformation
property of the auxiliary field $F_{\Phi}$ whose $vev$ breaks SUSY and
gives rise to gaugino masses.  We show the branching fraction for the
case where this field transforms as a singlet (dotted), a {\bf 24}
dimensional representation (solid) and a {\bf 200} dimensional
representation (dashed) of $SU(5)$ for {\it a})~$\mu> 0$, and {\it
b})~$\mu < 0$.  }
\label{fig:nunivm30}
\end{figure}
\newpage

\begin{figure}
\dofig{12cm}{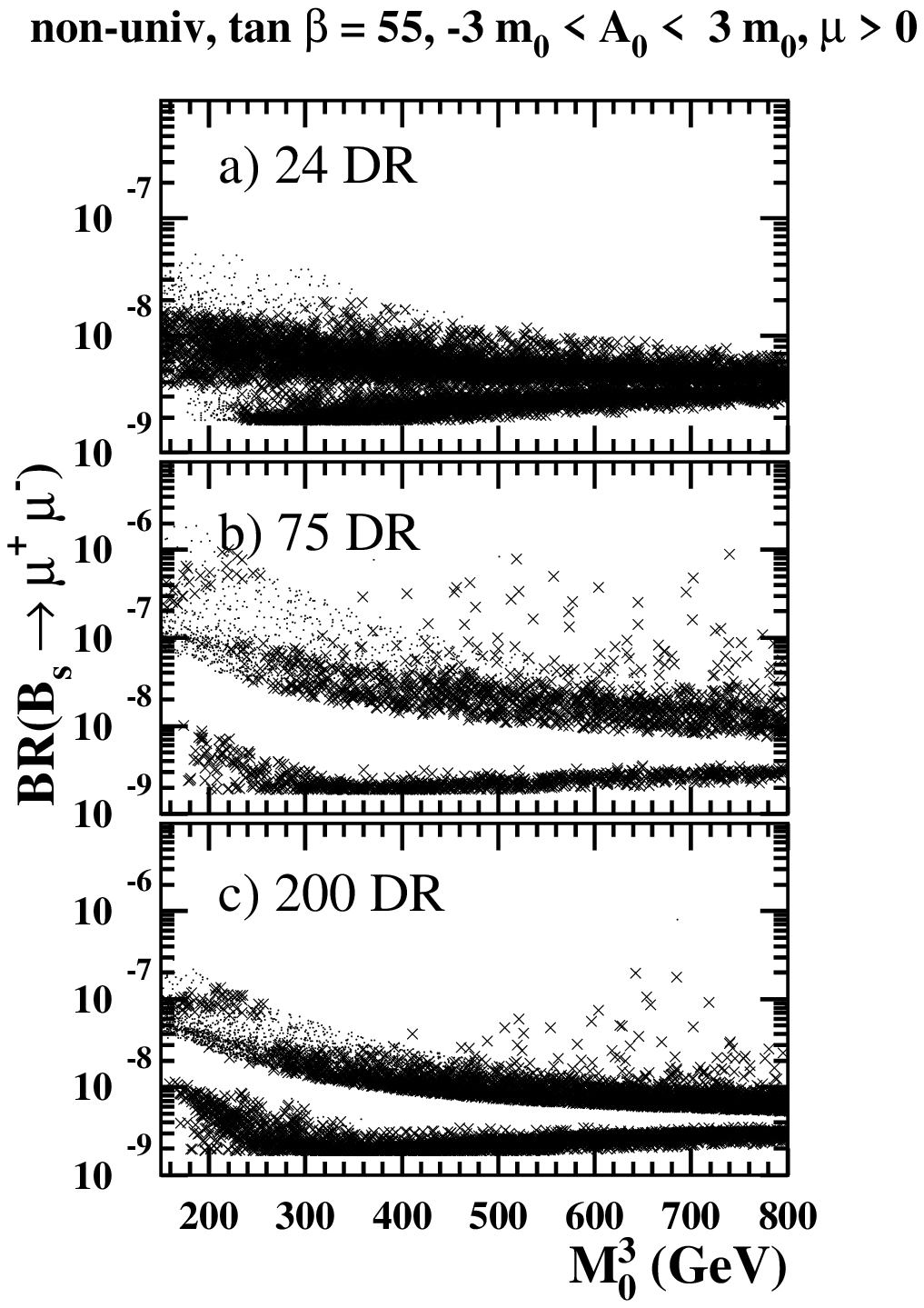}
\caption[]{The branching fraction for the decay $B_s \to \mu^+\mu^-$ for
a scan of the parameter space of $SU(5)$ supergravity models with
non-universal gaugino masses
over the range mentioned in the
text. We take $\mu > 0$. We show results versus the GUT scale gluino
mass parameter $M_3^0$ for the {\it a})~{\bf 24} model, 
{\it b})~{\bf 75} model, and {\it c})~{\bf 200} model discussed in the text.
Each cross (dot) denotes a model where $B(b\to s\gamma)$ lies within
(outside) the range $(2-5)\times 10^{-4}$.   }
\label{fig:nunivscanp}
\end{figure}
\newpage

\begin{figure}
\dofig{12cm}{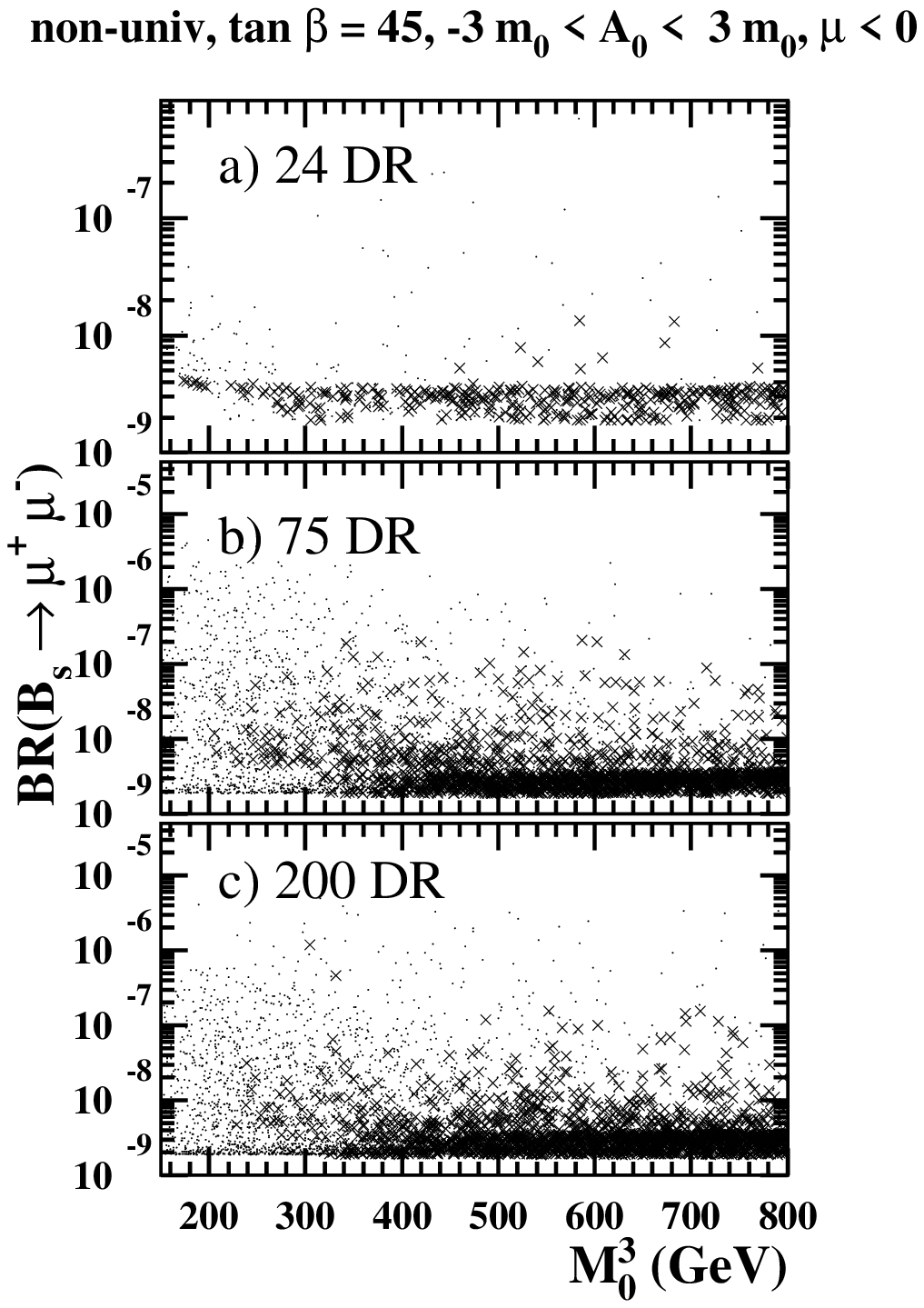}
\caption[]{ The same as Fig.~\ref{fig:nunivscanp}, except for $\mu < 0$.}
\label{fig:nunivscanm}
\end{figure}
\newpage

\end{document}